\documentclass[%
 reprint,
superscriptaddress,
 amsmath,amssymb,
 aps,
]{revtex4-1}

\usepackage{graphicx}
\usepackage{dcolumn}
\usepackage{bm}
\usepackage{hyperref}
\usepackage[capitalise]{cleveref}
\usepackage{siunitx}
\usepackage{wrapfig}
\usepackage{times,mathptmx}

\newcommand{\defeq}{\mathrel{\mathop:}=}

\begin{document}

\preprint{APS/123-QED}

\title{Elasticity of 3D networks with rigid filaments and compliant crosslinks}

\author{Knut M. Heidemann}
\affiliation{Institute for Numerical and Applied Mathematics, Georg-August-Universit{\"a}t, G{\"o}ttingen, Germany}
\author{Abhinav Sharma}%
\affiliation{Third Institute of Physics---Biophysics, Georg-August-Universit{\"a}t, G{\"o}ttingen, Germany}%

\author{Florian Rehfeldt}
\affiliation{Third Institute of Physics---Biophysics, Georg-August-Universit{\"a}t, G{\"o}ttingen, Germany}%

\author{Christoph F. Schmidt}
\email{cfs@physik3.gwdg.de}
\affiliation{Third Institute of Physics---Biophysics, Georg-August-Universit{\"a}t, G{\"o}ttingen, Germany}%

\author{Max Wardetzky}
\email{wardetzky@math.uni-goettingen.de}
\affiliation{Institute for Numerical and Applied Mathematics, Georg-August-Universit{\"a}t, G{\"o}ttingen, Germany}

\date{\today}



\begin{abstract}
Disordered filamentous networks with compliant crosslinks exhibit a low linear elastic shear modulus at small strains, but stiffen dramatically at high strains.
Experiments have shown that the elastic modulus can increase by up to three orders of magnitude while the networks withstand relatively large stresses without rupturing. Here, we perform an analytical and numerical study on model networks in three dimensions.
Our model consists of a collection of randomly oriented rigid filaments connected by flexible crosslinks that are modeled as wormlike chains.
Due to zero probability of filament intersection in three dimensions, our model networks are by construction prestressed in terms of initial tension in the crosslinks. We demonstrate how the linear elastic modulus can be related to the prestress in these network.
Under the assumption of affine deformations in the limit of \emph{infinite} crosslink density, we show analytically that the nonlinear elastic regime in 1- and 2-dimensional networks is characterized by power-law scaling of the elastic modulus with the stress.
In contrast, 3-dimensional networks show an exponential dependence of the modulus on stress.
Independent of dimensionality, if the crosslink density is \emph{finite}, we show that the only persistent scaling exponent is that of the single wormlike chain.
We further show that there is no qualitative change in the stiffening behavior of filamentous networks even if the filaments are bending-compliant.
Consequently, unlike suggested in prior work, the model system studied here cannot provide an explanation for the experimentally observed linear scaling of the modulus with the stress in filamentous networks.
\end{abstract}
\maketitle

\graphicspath{ {./img/} }

\section{Introduction}
The mechanical properties of biological cells are governed by the \emph{cytoskeleton}, a viscoelastic composite consisting of three main types of linear protein polymers: actin, microtubules, and intermediate filaments. These filamentous polymers are crosslinked by various binding proteins and constitute a dynamic complex network that maintains the structural integrity of the cell with the capacity for dynamic reorganization needed for active processes.
Many \textit{in vitro} studies have focused on reconstituted networks with rigid crosslinks \cite{Janmey1991,MacKintosh1997,Xu1998,Gardel2004,Gardel2004b,Storm2005,Bausch2006b,Koenderink2006,Chaudhuri2007,Janmey2007a,Kasza2007,Liu2007}. In the cytoskeleton, however, many of the crosslinks are themselves extended and highly compliant proteins.
Such flexible crosslinks can strongly affect the macroscopic network elasticity \cite{Gardel2006a,Gardel2006,DiDonna2007,Broedersz2008,Dalhaimer2007,Lee2009,Sharma2013a, Kasza2009, Kasza2010}.
Indeed, experimental studies show that composite networks can have a linear modulus as low as $\sim\SI{1}{\pascal}$, while being able to stiffen by up to a factor of \num{1000} \cite{Gardel2006,Kasza2007}.  

Here we analyze 3-dimensional (3D) composite networks theoretically, and we offer physical simulations thereof. 
Our networks are composed of randomly oriented rigid filaments
that are connected by highly flexible crosslinks, each of which is modeled as a wormlike chain (WLC) \cite{Bustamante1994,Marko1995a}, which has been shown to accurately describe flexible crosslinkers, such as filamin \cite{Schwaiger2004,Furuike2001}.
In our approach we assume that the filaments are much more rigid than the crosslinks, meaning that the network elasticity is dominated by the entropic stretching resistance of the crosslinks.

In our theoretical analysis we adopt the widely employed assumption of affine deformations \cite{Broedersz2008,Broedersz2009b,Sharma2013a}.
Under this premise, the network is assumed to deform affinely on the length scale of the filaments, which in turn is assumed to be much longer than the contour length of the crosslinks.
Using a single filament description in the limit of a \emph{continuous} distribution of crosslinks along the filament, we obtain the asymptotic scaling behavior of the elastic modulus with the stress in the nonlinear regime. We show that in 1-dimensional (1D) networks, the elastic modulus scales with the second power of the stress, whereas it scales with the third power in 2-dimensional (2D) networks.
Remarkably, there is no power law scaling in 3D---in fact, the elastic modulus of a 3D composite network increases exponentially with the stress.
Numerical evaluation of the affine theory at \emph{finite} crosslink densities---as opposed to a continuous distribution of crosslinks---shows that (i) the only asymptotic scaling is that of the modulus scaling with an exponent $3/2$ with the stress and that (ii) the dependence on dimensionality of the system is limited to an intermediate-stress regime. These findings are in agreement with our extensive \emph{physical simulations} of 3D composite networks. For all cases, the elastic modulus \emph{diverges} at a finite strain.

Our theoretical analysis is inspired by the mean-field model proposed by Broedersz et al. \cite{Broedersz2008,Broedersz2009b}. In sharp contrast to our theoretical analysis and to the results of our physical simulations, however, these authors predict \emph{linear} scaling of the elastic modulus with applied stress. In particular, for any finite strain, the elastic modulus remains \emph{finite} in their model. While this linear scaling of the elastic modulus is in accordance with what has been observed experimentally \cite{Gardel2006a,Kasza2009,Kasza2010}, we here argue that this model does not adequately capture the elastic response of networks with rigid filaments and 
\emph{permanent} (i.e., non rupturing or rebinding) crosslinks of finite length.

In Ref.~\cite{Sharma2013a}, the authors ruled out that the experimentally observed approximate linear scaling of the modulus with the stress might be be due to enthalpic (linear) \emph{stretching} compliance of the crosslinks or filaments. 
Here, we complement their analysis by physical simulations that take into account \emph{bending} of filaments. Our results empirically show that  
%
the inclusion of bending rigidity does not impact the nonlinear stiffening behavior of composite networks either.
We therefore conclude that the theoretical explanation for the linear scaling of the modulus with stress in experiments remains an challenging open problem.

By physical simulations, we also study the role of prestress. We show that in contrast to 1D and 2D networks, 3D networks experience an initial tension due to non-intersecting filaments resulting in initially stretched crosslinks, and are therefore prestressed.
The modulus in the linear deformation regime is then governed by this prestress; indeed, it is higher than the modulus expected from the affine theory. Our simulations additionally indicate that if the network is allowed to relax initial tension by unbinding and rebinding of crosslinks, the impact of prestress on the elastic modulus in the linear regime becomes insignificant,
although the prestress does not relax all the way to zero. 

The remainder of the article is organized as follows.
First, we present the affine theory of composite networks in \cref{sec:Affine theory}.
Under the assumption that deformations of the network are affine on the length scale of the filaments, we derive expressions for the stress and modulus in 1D, 2D, and 3D.
We then present our physical simulation model and describe our network generation procedure in \cref{sec:Simulation model}. 
We expand on the implications of our 3D simulation procedure in \cref{secImplications}; in particular, we explain the emergence of prestress.
We then discuss the results of our simulations in the linear deformation regime in \cref{sec:Linear regime} and indicate which results from the affine theory are still valid. 
Finally, we analyze the simulation results in the nonlinear regime in~\cref{sec:NonlinearRegime}. 
\section{Theory}
\label{sec:Affine theory}
In this section we analytically derive the stress and modulus of a composite network under the assumption of affine deformations on the length scale of the filaments.
We consider a collection of $N$ rigid filaments of length $L$ that are permanently connected by $nN/2$ flexible crosslinks of contour length $l_0$, where $n$ is referred to as the crosslink density, i.e., the number of crosslinks per filament.
The filaments are assumed to be perfectly rigid, i.e., they neither bend nor stretch, and the crosslinks are modeled via the WLC interpolation formula \cite{Marko1995a}
\begin{align}
  f_\text{cl}(u) = \frac{k_\text{B}T}{l_\text{p}}\left(\frac{1}{4(1-\frac{u}{l_0})^2}-\frac{1}{4}+\frac{u}{l_0}\right)\ ,
  \label{eq:wlcforce}
\end{align}
where $k_\text{B} T$ is the thermal energy, $l_\text{p}$ the persistence length and $u\geq0$ the end-to-end distance of the crosslink.
Assuming $l_0 \gg l_\text{p}$ this force-extension relation implements a crosslink rest-length of zero and shows a characteristic stiffening with divergence of force as $u \to l_0$. 
\Cref{eq:wlcforce} can be integrated to yield the energy \footnote[2]{More precisely, it is a \emph{free} energy, which includes both, energetic (bending) and entropic terms for the crosslinks (not for the filaments).} (up to a constant)
\begin{align}
 E_\text{cl}(u) = \frac{k_\text{B}T}{l_\text{p}}\left(\frac{l_0}{4(1-\frac{u}{l_0})}-\frac{l_0}{4}- \frac{u}{4} +\frac{u^2}{2 l_0} \right)\ .
  \label{eq:wlcenergy}
\end{align}

Imposing affine deformations on the filament level fully determines the deformation field $u$ on the subfilament level.
In the following analysis, we consider a single representative filament subject to an extensional strain of the surrounding medium that it is embedded in and crosslinked to. 
\subsection{1D network calculation}
\label{sub:1D}
We start with a one dimensional system, i.e., 1D extensional strain $\epsilon$, and advance in dimensionality by converting an applied shear strain $\gamma$ to the orientation dependent extensional strain $\epsilon(\gamma)$ felt by the filament.

\begin{figure}
  \centering
  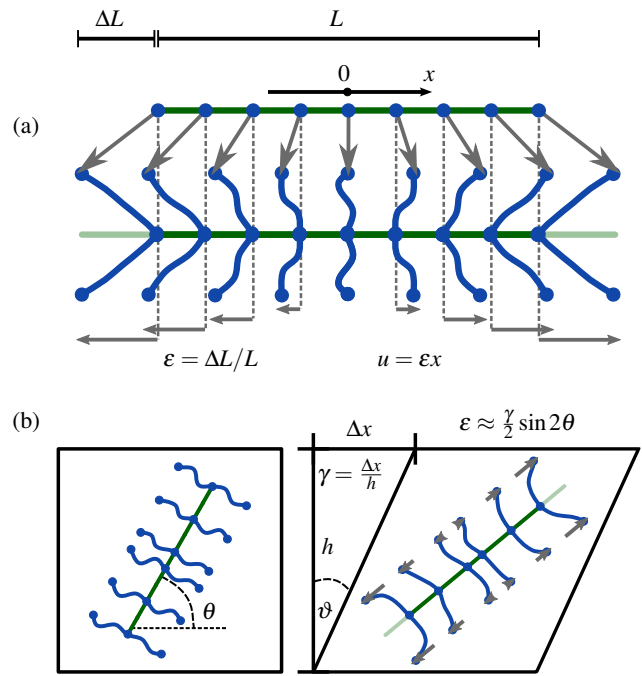
  \caption{Sketch of the assumptions of the affine theory: (a) 1D: A filament (green) of length $L$ is connected to its surrounding through $n$ crosslinks (blue) that have zero extension at zero strain. The surrounding of the filament is subject to a uniform extensional strain $\epsilon=\Delta L /L$. Since the filament itself is assumed to be perfectly rigid, all deformation goes into the crosslinks (drawn in y-direction for better visualization). The deformation of a crosslink at distance $x$ from the center of the filament is given by $u=\epsilon x$ (deformation field depicted by the horizontal gray arrows).
  (b) For a 2D system, the extensional strain on a filament at angle $\theta$ with the axis in shear direction is given by $\epsilon \approx (\gamma/2) \sin 2\theta$, for a small shear strain $\gamma=\Delta x/h=\tan{\vartheta}$.}
  \label{fighairyrod}
\end{figure}
In the rest frame of the filament, the end-to-end distance of a crosslink at distance $x$ from the center of the filament is given by $|u(x,\epsilon)|=|\epsilon x|$ (see~\cref{fighairyrod} (a)). 
For notational convenience, we consider positive $\epsilon$ only.
Under the assumption that the crosslink density is high enough that one can consider the associated distribution as uniformly continuous, the total energy of a filament in 1D is given by
\begin{align}
  E_\text{1D}(\epsilon) = 2 \frac{n}{L} \int_0^{L/2} E_\text{cl}(\epsilon x)\, dx\ .
  \label{eqEinfinite}
\end{align}
Substituting \cref{eq:wlcenergy} into \cref{eqEinfinite}, this expression can be integrated analytically (see~\cref{sub1Dappendix}).

Following the described approach for the linear regime of the WLC force-extension relation, i.e., for $u \ll l_0$, the linear modulus may be extracted as $G_0^\text{aff}=\tfrac{2E}{V \epsilon^2}$, where $E/V$ is the energy per unit volume $V$ in the network and $\epsilon$ is a small strain \cite{Landau1975elasticity}.
For a 1D system this yields $G_0^\text{aff}=\tfrac{1}{8} \rho n k_\text{cl}L$, with $k_\text{cl} = \tfrac{3}{2} \tfrac{k_\text{B} T}{l_\text{p} l_0}$ being the linear spring constant of a crosslink and $\rho\defeq NL/V$ the total length of filaments per unit volume. The same holds for the modulus in 2D and 3D, but with different numerical prefactors: $1/96$ and $1/192$, respectively \cite{Broedersz2008,Broedersz2009b,Sharma2013a}.

Next we show that one can extract a functional relation between nonlinear modulus and stress in the nonlinear regime, based on simple asymptotic scaling analysis. It follows from above that there is a strain $\epsilon_\text{d}\defeq l_0/(L/2)$ at which the outer most crosslink (at $x=L/2$) reaches maximum extension. For $\epsilon \to \epsilon_\text{d}$ the energy diverges as
\begin{align}
  E_\text{1D}^{\text{div}}(\epsilon) \sim -\frac{1}{\epsilon} \ln \left(1-\frac{\epsilon }{\epsilon_\text{d}}\right)\ ,
  \label{eq:E1D}
\end{align}
with `$\sim$' defined via $E\sim f \Leftrightarrow E/f \to \text{const.}$ The upper index `div' always indicates that we are only taking into account the diverging part of the 1D filament energy. 
We express stress and differential elastic modulus via $\sigma = \tfrac{1}{V} \tfrac{dE}{d\epsilon}$ and $K = \tfrac{1}{V} \tfrac{d^2 E}{d\epsilon^2}$, respectively, in order to obtain $\sigma_\text{1D} \sim 1/(1-\epsilon/\epsilon_\text{d})$, and $K_\text{1D} \sim 1/(1-\epsilon/\epsilon_\text{d})^{2}$.
We arrive at the asymptotic scaling relation
\begin{align}
  K_\text{1D} \sim (\sigma_\text{1D})^2\ .
\end{align}
This scaling relation between modulus and stress in 1D has also been derived in previous work \cite{Sharma2013a}. Next we consider scaling relations in 2D and 3D.
\subsection{2D network calculation} 
\label{sub:2D}
We perform similar calculations as in 1D, while taking into account that the extensional strain $\epsilon$, which results from a shear strain $\gamma$ on a 2D system, depends on the orientation of the filament under consideration.
In the small-strain limit one thus obtains
\begin{align}
  |\epsilon(\gamma,\theta)|=|(\gamma/2) \sin 2\theta|\ ,
\end{align}
where $\theta \in [0,\pi]$ is the angle between the filament and the shear direction (see~\cref{fighairyrod}(b)).

Substituting this expression into \cref{eq:E1D} and averaging over all orientations leads to
\begin{align}
  \langle E_\text{2D}^\text{div}\rangle_\theta(\gamma) \sim \int_0^{\pi/2} \frac{-\ln (1-\frac{\gamma L}{4l_0}\sin 2\theta )}{(\gamma/2) \sin 2\theta }\, d\theta\ ,
 \label{eq:E2Dint}
\end{align}
where we assume $\gamma\geq 0$ for notational convenience; the upper integration limit is reduced to $\pi/2$ because $|\sin 2\theta|$ is $\pi/2$-periodic. Note that we do not take into account a redistribution of filament orientations under the shear transformation.
This approach, as well as the small-strain approximation for $\epsilon(\gamma,\theta)$, are justified if $L\gg l_0$, since then the strain $\gamma_\text{d} \defeq  4l_0/L$ at which the integrand diverges is small.   

Differentiating \cref{eq:E2Dint} with respect to $\gamma$ and neglecting the weaker (logarithmically) diverging part of the integrand leads to an expression for the stress, as $\gamma \to \gamma_\text{d}$:
\begin{align}
  \langle\sigma_\text{2D} \rangle_\theta (\gamma) &\sim \int_0^{\pi/2}\frac{d\theta}{1-\left(\frac{\gamma}{\gamma_\text{d}}\right)\sin2 \theta}\ ,\\
 &= \frac{\pi-\arccos(1-\gamma/\gamma_\text{d})}{\sqrt{1-(\gamma/\gamma_\text{d})^2}}\ .
  \label{eq:2Dstress}
\end{align}
The divergence of the stress is of the form $\sigma_\text{2D} \sim 1/(1-(\gamma/\gamma_\text{d}))^{1/2}$ and hence $K_\text{2D} \sim 1/(1-\gamma/\gamma_\text{d})^{3/2}$.
Therefore, the asymptotic scaling behavior for the differential modulus in two dimensions is given by
\begin{align}
  K_\text{2D} \sim (\sigma_\text{2D})^3\ .
\end{align}
Note the difference of the scaling relations to the ones in the 1D scenario. Stress shows a weaker divergence with strain than in 1D but a stronger dependence on the differential modulus. 
Integration of the diverging part of the stress further shows that the energy at maximum strain is finite---in contrast to the 1D setting, where the energy diverges at the critical strain. This is an effect introduced by orientational averaging only.
\subsection{3D network calculation} 
\label{sub:3D}
For a 3D network, the extensional strain on a filament in the small-strain limit is given by
\begin{align}
  |\epsilon(\gamma,\theta,\phi)| = |(\gamma/2) \sin2\theta \cos\phi| \ ,
\end{align}
in the usual spherical coordinates.
In direct analogy to the 2D case (see \cref{eq:2Dstress}), the averaged stress close to $\gamma_\text{d} = 4l_0/L$ can be written as
\begin{align}
  \langle \sigma_\text{3D}\rangle_{\theta,\phi}(\gamma) \sim \int\limits_0^{\pi/2}\int\limits_0^{\pi/2} \frac{\sin \theta\, d\phi d\theta}{1-\left(\frac{\gamma}{\gamma_\text{d}}\right)\sin 2\theta \cos \phi} \ ,
  \label{eq:3Dstress}
\end{align}
with $\gamma \geq 0$; the upper integration limit for the $\phi$ integration is reduced to $\pi/2$ because $|\cos \phi|$ is $\pi$-periodic and symmetric about $\pi/2$ on $[0,\pi]$.
If we carry out the $\phi$ integral and expand the integrand around $\theta = \pi/4$, in order to integrate over $\theta$ (see~\cref{sub:3D network} for details), we obtain 
  $\sigma_\text{3D} \sim -\ln(1-\gamma/\gamma_\text{d})$ and hence $K \sim 1/(1-\gamma/\gamma_\text{d})$.
Consequently, $K$ does not scale with $\sigma$ as a power law; instead, one obtains
\begin{align}
  K_\text{3D} \sim e^{c  \sigma_\text{3D}} \ ,
\end{align}
with a real constant $c$.
The absence of asymptotic power law scaling sets 3D networks apart from 1D and 2D networks. In 3D, we observe the weakest (logarithmic) divergence of stress with strain. 
Integrating the diverging part of the stress shows that the energy again remains finite for $\gamma \to \gamma_\text{d}$.

\begin{figure}
  \centering
  \includegraphics{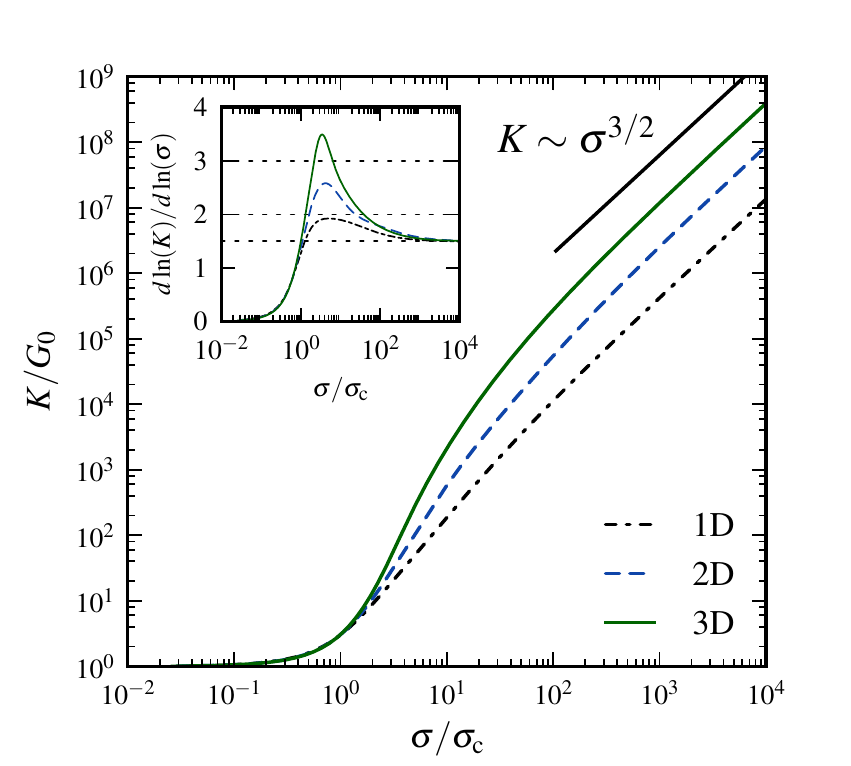}
  \caption{Differential modulus $K$ as a function of shear stress $\sigma$ in the affine limit, with finite number of crosslinks ($n=60$), rescaled by the linear elastic modulus $G_0\defeq K|_{\gamma=0}$ and critical stress $\sigma_\text{c}\defeq \sigma(\gamma_\text{c})$, respectively, where $\gamma_\text{c}$ is defined via $K(\gamma_\text{c})=3G_0$. Straight line indicates power law scaling $K\sim \sigma^{3/2}$. Inset shows local slope $d \ln K / d \ln \sigma$; dotted lines indicate power law scaling with exponents from affine theory $\{2,3\}$ and single WLC scaling $\{3/2\}$.
  Independent of dimensionality, the asymptotic large stress scaling is $K \sim \sigma^{3/2}$. In an intermediate-stress regime, the theoretical values for infinite crosslink densities are approached.}
  \label{figkvssigmatheory}
\end{figure}

\paragraph*{Finite crosslink density.}~By considering the limit of infinite crosslink density, we have derived theoretical scaling relations for strain stiffening by integrating along a filament's backbone (see~\cref{eqEinfinite}).
For any real system, however, the crosslink density is finite and \cref{eqEinfinite} turns into a sum
\begin{align}\label{eq:finite_sum}
  E = \sum_{i=1}^n E_\text{cl}(\epsilon x_i) \ ,
\end{align}
where $\{x_i\}$ are the crosslink binding sites along the filament. \cref{figkvssigmatheory} shows numerical results for the behavior of the corresponding differential modulus $K$ for finite $n$, obtained by numerical evaluation of \cref{eq:finite_sum} and proper orientational averaging.
Note that the asymptotic scaling behavior of $K$ 
in the limit of infinite crosslink density influences a finite network's behavior 
in the \emph{intermediate}-stress regime (see~inset of~\cref{figkvssigmatheory}); however, near the critical strain, the differential modulus scales as $K\sim\sigma^{3/2}$, i.e., like the response of a single WLC. Furthermore, for 1D and 2D systems the theoretical scaling exponents in the limit of infinite crosslink densities can (in the intermediate regime) indeed be approached by increasing $n$.  In contrast, as shown above, in 3D the theoretically derived scaling of $K$ is exponential in $\sigma$. Such an exponential increase is quantified by an (in principle) indefinitely increasing maximal slope  with increasing $n$ in the $\ln K$ versus $\ln \sigma$ plots; e.g., for $n=\num{60}$ the maximal slope is \num{3.49}, for $n=\num{3000}$ it is \num{5.82}.
However, for any \emph{finite} $n$, eventually there is always a universal scaling of $K \sim \sigma^{3/2}$, resulting from the single WLC force-extension relation, independent of the dimensionality of the system. Indeed, for any given $n$, the integral representation \cref{eqEinfinite} becomes invalid close to $\gamma=\gamma_\text{d}$ due to the divergence of the WLC energy.

The numerical results in \cref{figkvssigmatheory} have been obtained without the small-strain approximation for the extension of the filaments. However, redistribution of the filament orientations under shear has not been taken into account in \cref{figkvssigmatheory}. Calculations including this effect show that it 
may both decrease 
and increase 
the maximum intermediate slope in the $\ln K$ versus $\ln \sigma$ plot and shift the peak to larger stress values depending on the maximum strain $\gamma_\text{d}$. In any case, the asymptotic scaling regime remains unchanged.

In the next section we introduce the simulation framework that we use to study 3D networks consisting of many filaments and crosslinks, relaxing the assumption of affine deformations.
\section{Simulation model} 
\label{sec:Simulation model}
We perform quasistatic simulations of 3D networks that consist of $N$ rigid filaments of length $L$, permanently crosslinked by a collection of $nN/2$ crosslinks of length $l_0$.
All lengths are measured in units of the side length of the cubic periodic simulation box.
A typical set of parameters is $N=3000$, $L=0.3$, $n=60$, $l_0=0.03$.

Each filament is modeled as perfectly rigid, implying that its configuration can be described by its two endpoints only, which are constraint to stay at distance $L$. The flexible crosslinks are modeled as a central force acting between the two binding sites. 
In particular, we use the WLC interpolation formula (\cref{eq:wlcforce}) and the corresponding energy (\cref{eq:wlcenergy}).
In all data that is presented, forces are measured in units of $(k_\text{B}T)/l_\text{p}$.
There are no additional degrees of freedom introduced through the crosslinks, since their configuration is represented via the endpoints of the filaments, in terms of barycentric coordinates.

\begin{figure}
  \centering
  \includegraphics[width=\columnwidth,trim=3.5cm 4.0cm 3.5cm 4.0cm,clip=true]{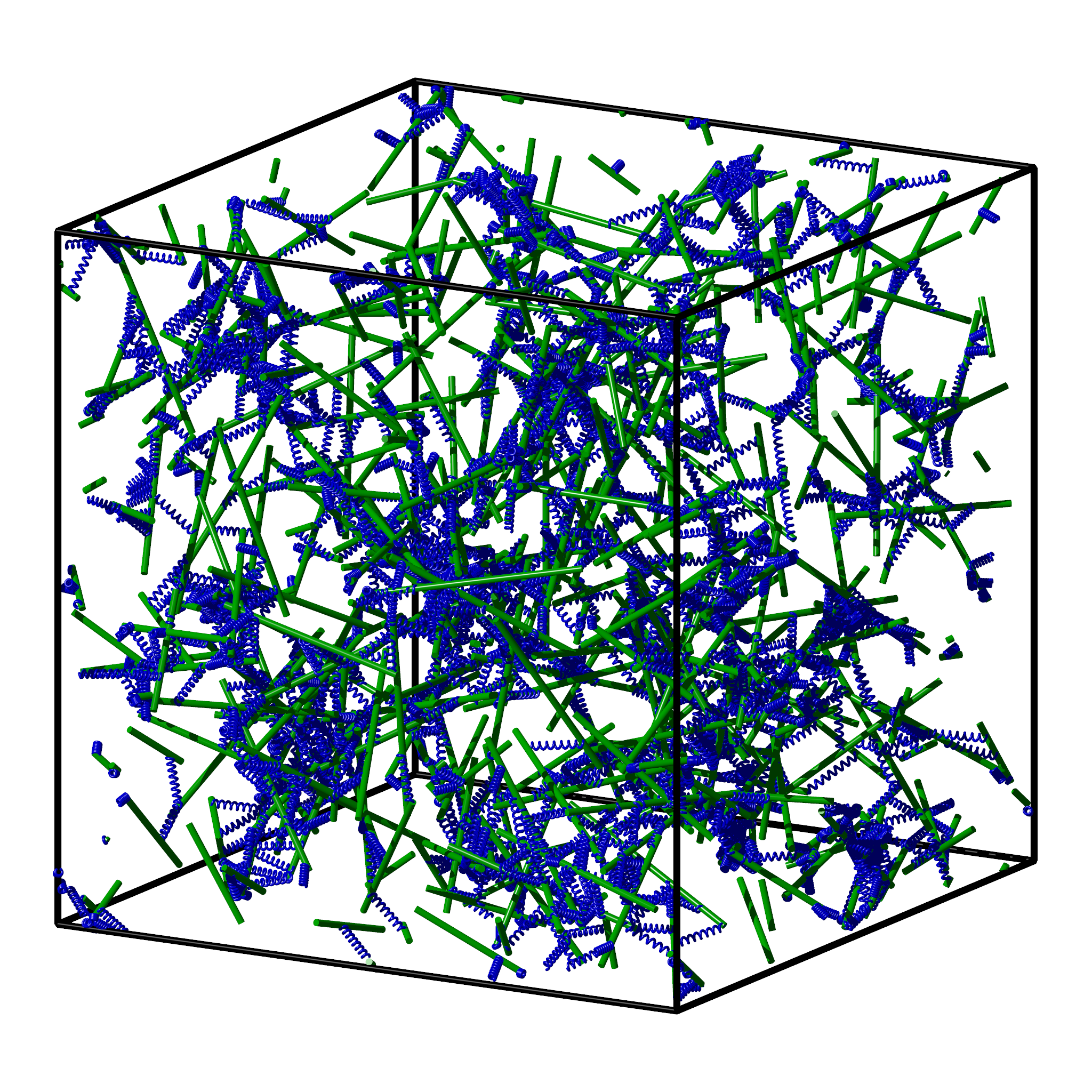}
  \caption{Example of an initially generated network that has not been relaxed into static equilibrium yet. Rigid filaments are shown in green, flexible crosslinks in blue. Short crosslink or filament fragments correspond to filaments/crosslinks that cross the periodic boundaries of the simulation box.
  For the sake of visual appearance, the network is much sparser than the systems that are studied in the remainder of this article, and the ratio of filament to crosslink length is much smaller, $N=300$, $n=10$, $L=0.3$, $l_0=0.1$, $\alpha=0.9$.\label{figSimbox}}
\end{figure}

In order to generate an initial network configuration we proceed as follows. We generate $N$ randomly distributed filaments by first randomly choosing their centers of mass in our simulation box and by then picking a random orientation for each filament.
For crosslinking we apply the following iterative procedure. We randomly choose two points on distinct filaments and insert a crosslink if the corresponding point-to-point distance is shorter than a certain threshold $\alpha l_0$. Here $\alpha \in [0,1)$
serves as a parameter to vary the initially allowed crosslink lengths in the system.
This procedure is repeated until the desired number of crosslinks is reached; see~\cref{figSimbox} for an illustration of the final configuration.
Since we perform quasistatic simulations, the system must be at static equilibrium at all times. As practically all crosslinks will be stretched beyond their rest-length after the initial network generation, we minimize the energy (of the crosslinks) before subjecting the simulation box to any deformation \footnote[3]{We do neither take into account fluctuations of the filaments nor excluded-volume effects.}.
For energy minimization we use the freely available external library IPOPT \cite{Wachter2006}, which requires the gradient and the Hessian of the system's energy function. 
It might happen during the optimization process, that individual crosslinks reach extensions $u$ larger than their contour length $l_0$. 
Acceptance of these solutions is prohibited by setting the energy to infinity ($10^{19}$) for $u\geq l_0$ in \cref{eq:wlcenergy}; without this modification it would become negative in that regime.
The length constraints for the filaments are realized via Lagrange multipliers.

In order to extract elastic properties of the network we perform quasistatic shearing by applying an affine incremental shear strain $\delta \gamma$ to the network, with subsequent rescaling of filaments to length $L$ (see~\cref{fighairyrod}). We apply Lees-Edwards shearing periodic boundary conditions \cite{Lees1972}.
The magnitude of $\delta\gamma$ is determined by calculating the maximum affine shear that leaves all crosslinks below their contour length. 
Due to the rescaling of filament lengths, a nonaffine deformation component is introduced. This nonaffinity may lead to crosslinks being overstretched after all. In this case, we iteratively halve the shear strain until the length of all crosslinks remains below their contour length.
After each shear increment, the energy is minimized. 
We apply a fixed upper bound of {\SI{1}{\percent} strain} on $\delta\gamma$ in order to stay reasonably close to the previous solution. This increases numerical efficiency and accelerates convergence because it allows us to use a warm-start procedure that reuses Lagrange multipliers from one minimization as initial guesses for the next one.
Moreover, the application of small shear steps reduces the likelihood of discontinuously jumping between local energy minima.

We stop shearing when the achievable increment in shear strain becomes smaller than a chosen threshold due to crosslinks that are very close to their maximum extension. During the entire simulation process, we record network parameters in the equilibrated states---in particular, the energy $E$ as a function of shear strain $\gamma$. This allows us to extract the shear stress $\sigma = \frac{1}{V} \frac{d E}{d\gamma}$ as well as the differential shear elastic modulus $K=\frac{d\sigma}{d\gamma} = \frac{1}{V}\frac{d^2E}{d\gamma}$.
Derivatives are taken by first interpolating $E(\gamma)$ with a cubic spline.
We define the linear shear elastic modulus as
\begin{align}
G_0\defeq K|_{\gamma=0} \ .
\label{eqDifferentialModulus}
\end{align}

In the following section we discuss the implications of our specific simulation model, in particular with respect to network structure, and contrast it with previous studies that have been carried out mostly in 2D.

\section{Initial tension and prestress}
\label{secImplications}
As mentioned in~\cref{sec:Simulation model}, our network generation results in a non-zero initial energy $E_0$ at zero strain. Indeed, by \emph{randomly} placing (zero-diameter) filaments in a 3D container, filaments have zero probability to intersect; thus, crosslinks have finite initial extension with probability one. 
This is different from 2D, where randomly placed filaments mutually intersect with a probability approaching one as their number increases. Indeed,
so-called Mikado models \cite{Sharma2013a,Wilhelm2003d,Head2003,Onck2005}, where filaments are crosslinked at their intersection sites \emph{only}, exhibit no forces at zero strain.

\begin{figure}
  \centering
  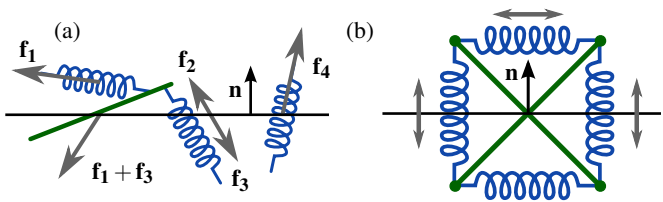
  \caption{(a) Measuring the total prestress $\sigma_0$ by extracting the normal component of the total force acting on a shear plane. We sum up all the forces acting on one side of the plane exerted by (i) the crosslinks passing through (here $\mathbf{f_2}$ and $\mathbf{f_4}$) and (ii) the filaments passing through (here $\mathbf{f_1}+\mathbf{f_3}$)---then we project onto the normal vector $\mathbf{n}$.
  (b) A tensegrity structure (here: \emph{Snelson's X} \cite{Connelly1998a}) remains in static equilibrium without application of boundary conditions. The forces acting on any plane add up to zero, i.e. no plane carries any total prestress although it is under tension locally. 
  \label{figStresstensor}}
\end{figure}

In contrast, the initial stretching of crosslinks in our networks results in an initial tension before any deformation. For a quantitative analysis we measure a global variant of this effect by what we call \emph{total prestress}
$\sigma_0$, which measures the \emph{normal stress} \footnote[4]{Note that our notion of prestress is not to be confused with the constant prestress externally applied in bulk rheology experiments, which is a \emph{shear stress} in general.} component orthogonal to the shear planes \footnote[5]{Although we could in principle define total prestress as the normal component of the stress acting on \emph{any} plane in our system we prefer to use shear planes as this simplifies the forthcoming analysis.}.
More precisely, we measure the single sided (e.g., upward) normal component of the force that is acting on a given shear plane, by summing up the normal components of the forces exerted by each crosslink and filament passing through the given shear plane, see~\cref{figStresstensor} (a).
The normal stress is then given by dividing by the surface area of the shear plane. Note that $\sigma_0$ does not depend on the choice of a particular shear plane; indeed, if the total stress was changing during vertical movement of a shear plane, then this would immediately contradict force balance in the system.

Intuitively, one might expect negative normal stresses (pulling down on the upper face of the simulation box), since crosslinks are \emph{contractile}.
However, since filaments withstand compression, it is possible to construct systems that exhibit positive normal stress.
This suggests the existence of configurations with zero normal stress \footnote[6]{Note that individual crosslinks are still under tension; however, the total normal force acting on the shear plane vanishes.}.
Indeed, so-called \emph{tensegrity structures} \cite{pugh1976introduction}, which are in static equilibrium in the absence of boundary conditions satisfy this criterion---while still being able to store arbitrary amounts of energy (see~\cref{figStresstensor} (b)).
Empirically, our simulations show that the random networks generated by the procedure described in \Cref{sec:Simulation model} exhibit negative initial normal stresses throughout. Their integrity is provided through the application of periodic boundary conditions. Note in particular, that our setup enforces conservation of volume of the simulation box. In general, it would be possible to relax the prestress by letting the volume of the simulation box change.
However, we did not follow this approach in the study presented here, in order to ensure that the filament length remains significantly smaller than the size of the simulation box.

In the following, we relate total prestress to the linear elastic response of our networks.
\section{Linear regime}
\label{sec:Linear regime}

In previous work \cite{Broedersz2008,Broedersz2009b,Sharma2013a}, an expression for the linear modulus in 3D was derived under the assumption of affine deformations and in absence of any initial tension in the network. Our simulations show that the linear elastic modulus depends on the initial tension in the network. 

\begin{figure}
  \centering
  \includegraphics{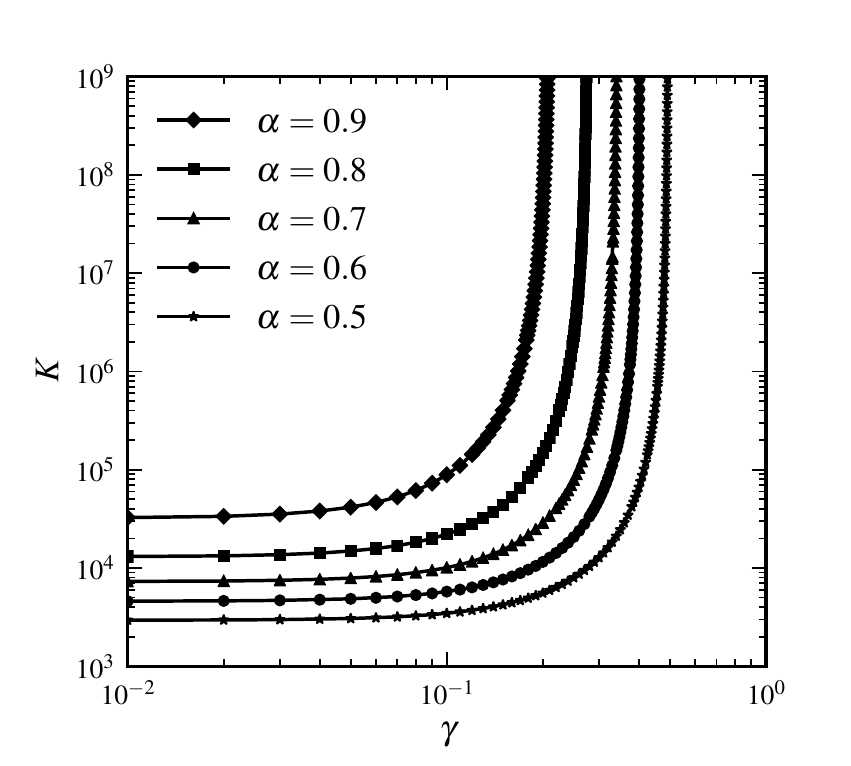}
  \caption{Differential elastic modulus $K$ as a function of strain $\gamma$ for different levels of initial tension. The initial tension in the network is varied by changing the initially admissible maximal crosslink length $\alpha l_0$. The linear modulus $G_0=K|_{\gamma=0}$ increases with the initial tension in the network (initial tension increases with $\alpha$). It is also evident that the divergence of $K$ occurs at a strain $\gamma_\text{d}$ that decreases with increasing $\alpha$. Here: $N=3000$, $n=60$, $L=0.3$, $l_0=0.03$.}
  \label{fig:KvsGamma-initialSpringLength}
\end{figure}

One scenario that clearly demonstrates the dependence of the linear modulus $G_0$ (defined in~\cref{eqDifferentialModulus}) on the initial tension is illustrated in \cref{fig:KvsGamma-initialSpringLength} where the admissible maximum initial crosslink length was varied.

\begin{figure}
  \centering
  \includegraphics{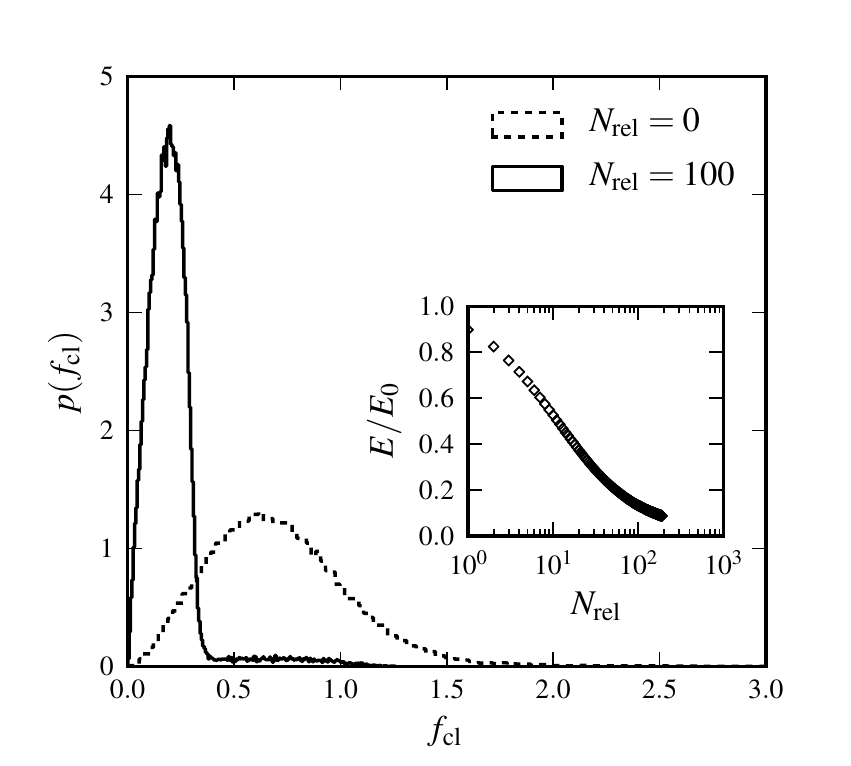}
  \caption{Distribution of forces in crosslinks for a system without or with $N_\text{rel}=100$ relaxation steps. The relaxation procedure cuts the large force tail of the initial distribution and establishes a sharper peak at small forces. The inset shows the total energy $E$ in the system, normalized by the initial energy $E_0$, as a function of number of relaxation steps $N_\text{rel}$.}
  \label{figforceDistributions}
\end{figure}

For a more quantitative analysis we have designed a method that allows us to change initial tension for a network with a fixed set of simulation parameters.
We first randomly generate a network as described above and let it relax into static equilibrium.
We then remove a given amount (\SI{5}{\percent}) of the most-stretched crosslinks in the system. Then we reconnect those crosslinks randomly again, and let the network relax.
This procedure is repeated $N_{\text{rel}}$ times.
Thereby, we successively decrease the system's initial tension, and therefore also its total energy, see~inset of~\cref{figforceDistributions}. Not only does the total energy decrease, we also observe a change in the distribution of forces (see~\cref{figforceDistributions}).
As long as one performs the crosslink binding-unbinding procedure over a small enough fraction of crosslinks, the network remains nearly isotropic.

\begin{figure}
\centering
\includegraphics{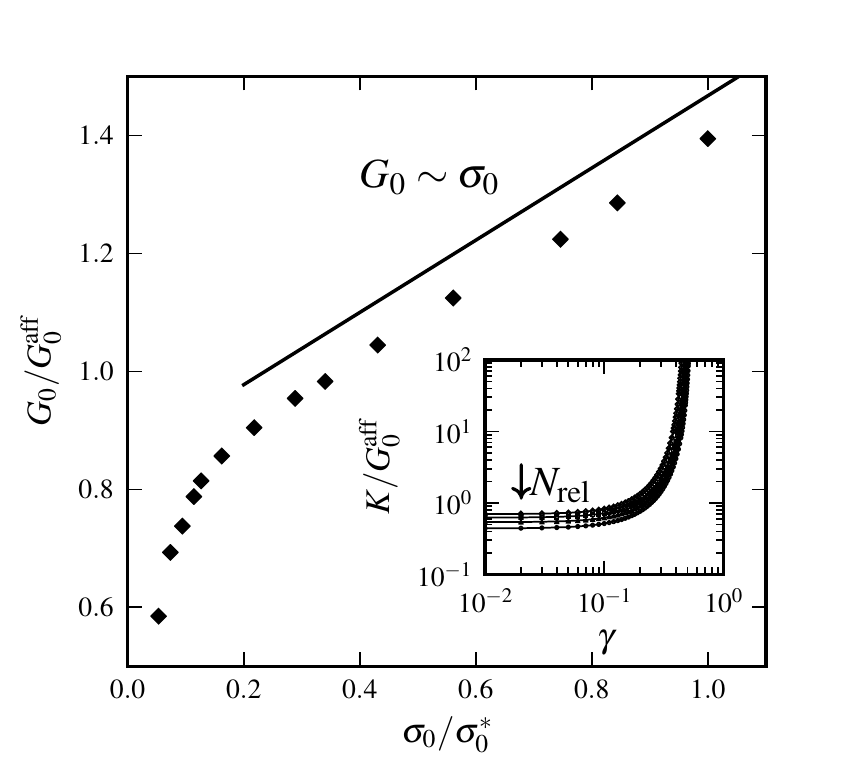}
\caption{Linear elastic modulus $G_0$ normalized by the affine prediction $G_0^\text{aff}$ as a function of total prestress $\sigma_0$ normalized by the total prestress $\sigma_0^*$ immediately after initial network generation. The total prestress is reduced via the procedure described in \cref{sec:Linear regime}.
For small total prestress, $G_0$ exhibits superlinear dependence on $\sigma_0$. Up to $\sigma_0=\sigma_0^*$, we observe linear scaling $G_0\propto \sigma_0$, as predicted by the model. The straight line is drawn as a guide to the eye, representing linear scaling. 
Parameters: $N=3000$, $n=60$, $L=0.3$, $l_0=0.06$, $\alpha=0.5$.
The inset shows differential elastic modulus $K$ versus shear strain $\gamma$ for systems with varying number of relaxation steps $N_\text{rel}\in\{0,50,100,150\}$. $G_0$ goes down with increasing $N_\text{rel}$.
Parameters: $N=3000$, $n=60$, $L=0.3$, $l_0=0.03$, $\alpha=0.5$.
}
\label{fig:G0vsPrestress-prestressRelaxation}
\end{figure}

It is apparent from the inset of \cref{fig:G0vsPrestress-prestressRelaxation} that the linear elastic modulus is reduced by increasing the number of relaxation steps, as expected. 
\cref{fig:G0vsPrestress-prestressRelaxation} also shows the dependence of linear modulus $G_0$ on the total prestress $\sigma_0$, which has been introduced in \cref{secImplications}.
We varied $\sigma_0$ via the above described procedure, and measured $G_0$ with the shearing protocol described in \cref{sec:Simulation model}.
After a certain number of relaxation steps the empirical value for $G_0$ equals the value $G_0^\text{aff}$ expected from affine theory (see~\cref{sub:1D}). Relaxing initial tension further, we reach moduli even below $G_0^\text{aff}$.
This is possible because the network can rearrange nonaffinely, thereby softening its response.
Over a certain range of total prestresses, we observe linear scaling of $G_0$ with $\sigma_0$, a phenomenon, which has been discussed in other contexts before (see for example Ref.~\cite{Alexander1998}).
%
\begin{figure}
  \centering
  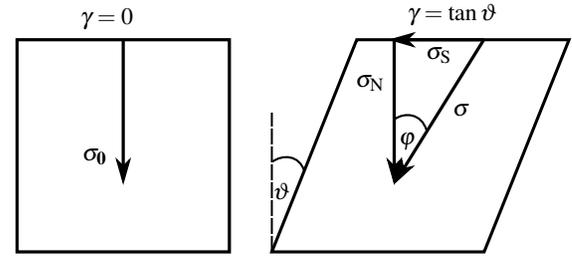
  \caption{
  The initial network carries a total prestress $\sigma_0$. After a small shear $\gamma=\tan \vartheta$ has been applied it exhibits a shear stress $\sigma_\text{S}$ and normal stress $\sigma_\text{N}$, with $\tan \varphi = \sigma_\text{S}/\sigma_\text{N}$. 
  \label{figSingleSpringModel}}
\end{figure}
We explain the linear regime as follows.
For small strains the normal component $\sigma_\text{N}$ of the stress acting on shear planes is close in magnitude to the total prestress $\sigma_0$, 
%
i.e., $\sigma_\text{N} \approx \sigma_0$.
For small strains given by shear angles $\vartheta \approx 0$, total forces acting on the shear planes make an angle $\varphi$ with the direction normal to the shear planes (see~\cref{figSingleSpringModel}). Our simulations show that $\tan \varphi \propto \tan \vartheta$ and that the constant of proportionality remains unchanged in the linear scaling regime. 
Therefore, shear satisfies
\begin{align}
  \gamma = \tan \vartheta \propto \frac{\sigma_\text{S}}{\sigma_{0}} \ ,
  \label{eqsinglespring}
\end{align}
where $\sigma_\text{S}$ is the component of the stress acting on shear planes in the shear direction, see~\cref{figSingleSpringModel}.
Hence, the linear elastic shear modulus $G_0$ defined via $\sigma_\text{S}=G_0 \gamma$ is proportional to the total prestress $\sigma_0$ via \cref{eqsinglespring}.
However, for very small total prestresses, i.e., after many relaxation steps, the modulus shows a steeper than linear dependence on $\sigma_0$. Indeed, in this regime the aforementioned constant of proportionality becomes larger. This effect might be attributed to the fact that for small $\sigma_0$, tensegrity type elements (see~\cref{figStresstensor} (b)), which do not contribute to the total prestress but carry energy, contribute significantly to the measured shear stress, thereby increasing $\varphi$ (see \cref{figSingleSpringModel}).

\begin{figure}
  \centering
  \includegraphics{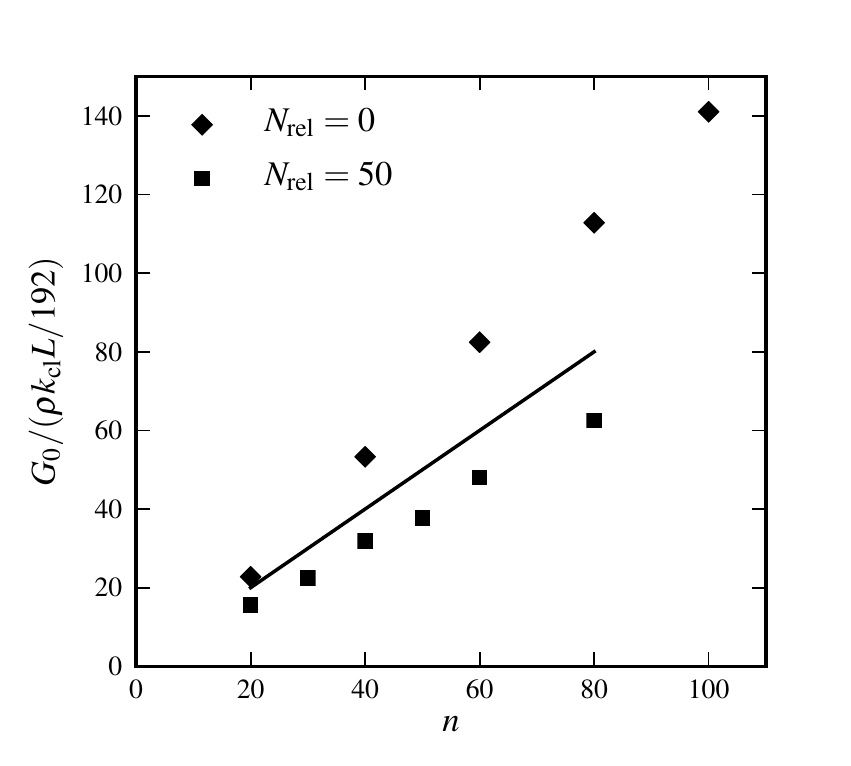}
  \caption{Linear elastic modulus $G_0$ versus crosslink density $n$ for systems with different number of relaxation steps: $N_\text{rel}=0$ (diamonds) and $N_\text{rel}=50$ (squares). Solid line indicates values expected from affine theory: $G_0^\text{aff}=\rho n k_\text{cl} L /192$. Parameters: $N=3000$, $L=0.3$, $l_0=0.06$, $\alpha=0.5$.}
  \label{figG0vsCrosslinkDensity}
\end{figure}
Furthermore, affine theory predicts linear scaling of the modulus $G_0$ with crosslink density $n$. \cref{figG0vsCrosslinkDensity} shows that this linear scaling is indeed reproduced in our simulations, independent of the prestress.
Moreover, by changing the prestress via our relaxation procedure it is possible to reach comparable slopes to what is predicted by the affine theory.

The next section deals with the nonlinear elastic response of the simulated networks, and relates it to the theoretical results that were derived in \cref{sec:Affine theory}.
\section{Nonlinear regime}
\label{sec:NonlinearRegime}
\subsection{Critical strain}
The networks that we study are inherently nonlinear because crosslinks are WLCs with finite length $l_0$ (see~\cref{eq:wlcforce}), resulting in pronounced strain stiffening at a critical strain $\gamma_\text{c}$. Stress diverges at a higher strain $\gamma_\text{d}$.
In our simulations, we define the critical strain $\gamma_\text{c}$ to be the strain where $K/G_0\approx 3$.
In the affine theory, $\gamma_\text{d}$ and $\gamma_\text{c}$ scale linearly with the ratio of crosslink to filament length $l_0/L$.
In our simulations, we cannot conclusively report on this dependence because 
the accessible ranges for $l_0$ and $L$ are quite limited. On the one hand, there exists an upper limit for $L$ (therefore also for $l_0$, since $l_0/L \ll 1$ should hold) to be significantly smaller than the simulation box. On the other hand, $L$ and $l_0$ are bounded from below due to computational limitations---this is because we need to increase the number of filaments in order to keep networks homogenous. 

\begin{figure}
  \centering
  \includegraphics{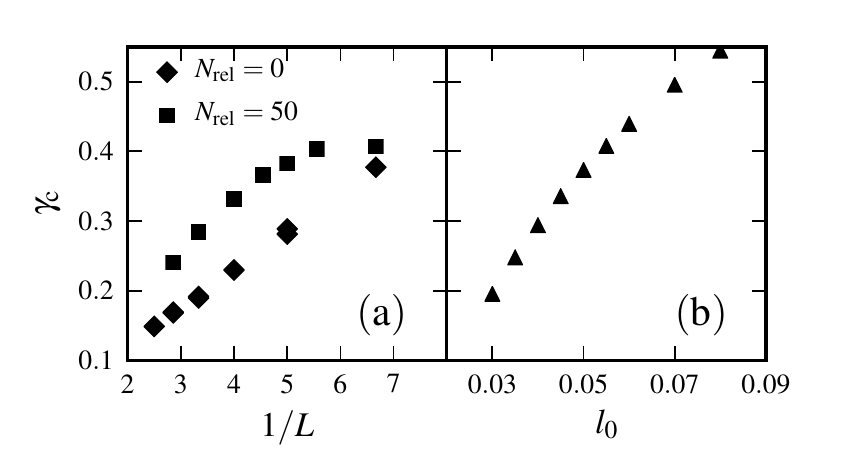}
  \caption{(a) Critical strain $\gamma_\text{c}$ versus inverse filament length $1/L$ for $N_\text{rel} =0$ and $N_\text{rel}=50$. Other parameters: $N=5000$, $n=60$, $l_0=0.04$, $\alpha=0.7$. We observe linear scaling $\gamma_\text{c} \propto 1/L$ for $N_\text{rel}=0$; systems in which relaxation has been applied show deviations from this behavior (see~$N_\text{rel}=50$ here). (b) Critical strain $\gamma_\text{c}$ versus crosslink contour length $l_0$ for a system with $N=3000$, $n=50$, $L=0.3$, $\alpha=0.5$.
  \label{figgammaCriticalvsL}}
\end{figure}
For ranges that are accessible to our simulations, we obtain the following results.
If we fix $l_0$, then we observe linear scaling $\gamma_\text{c} \propto 1/L$ for systems where no relaxation procedure has been applied (see~\cref{figgammaCriticalvsL} (a)).
Relaxed systems, however, sometimes show a less than linear dependence. This effect might be due to anisotropies induced by the relaxation procedure. 
If we fix $L$, then the dependence of $\gamma_\text{c}$ on $l_0$ is slightly less than linear (see~\cref{figgammaCriticalvsL} (b)).
\subsection{Differential modulus}
\begin{figure}
  \centering
  \includegraphics{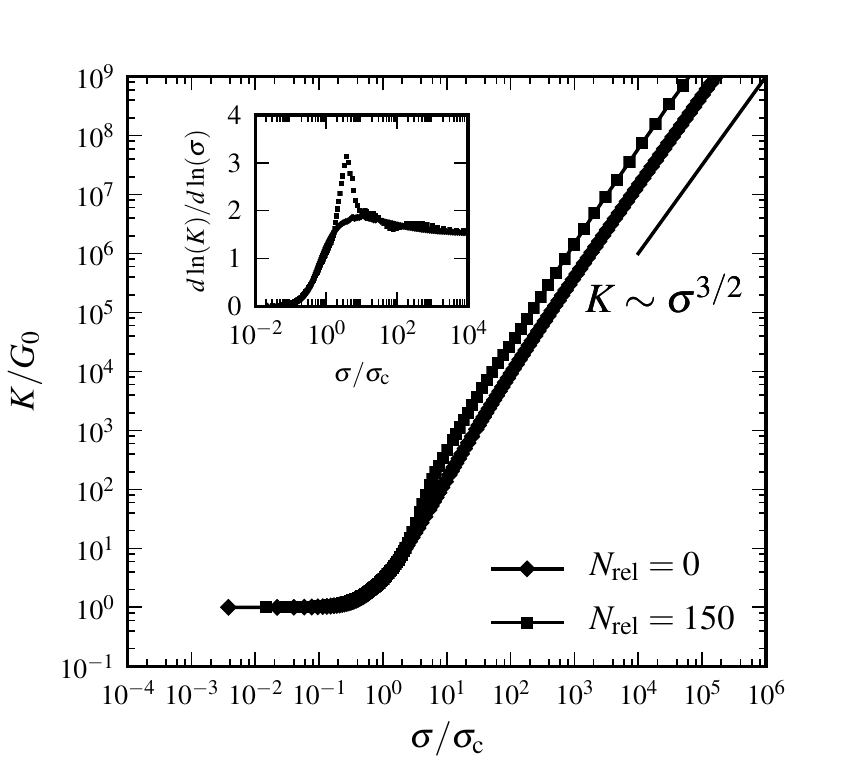}
  \caption{Differential modulus as a function of shear stress, rescaled by linear modulus and critical stress $\sigma_\text{c}=\sigma(\gamma_\text{c})$, respectively. Parameters: $N=3000$, $n=60$, $L=0.3$, $l_0=0.06$, $\alpha=0.5$, with ($N_\text{rel}=150$) and without ($N_\text{rel}=0$) relaxation. Inset shows the local slope $d \ln (K) / d \ln (\sigma)$ from the main plot. For large stresses, we observe power law scaling $K \sim \sigma^{3/2}$ (solid straight line). For intermediate stresses we recover slopes in the range of those derived from affine theory.}
  \label{fig:KvsSigma-initialSpringLength-rescaled}
\end{figure}
It remains to discuss the dependence of the differential modulus on stress, the affine theory of which has been derived in \cref{sec:Affine theory}.
For finite crosslink densities, the only persistent scaling behavior is $K\sim \sigma^{3/2}$, as $\gamma$ approaches $\gamma_\text{d}$---due to the fact that eventually single WLC response dominates.
In an intermediate regime, above the critical stress $\sigma_\text{c}=\sigma(\gamma_\text{c})$, we observe slopes $(d\ln K/d\ln \sigma)>3/2$. 
The majority of the simulations shows intermediate slopes around 2 or slightly above, mostly independent of simulation parameters, but there are also realizations that show maximum slopes up to 3.5 (see~\cref{fig:KvsSigma-initialSpringLength-rescaled}).
These higher slopes and the final scaling $K \sim \sigma^{3/2}$ are in accordance with the predictions of affine theory. Indeed, a slope of 3.5 is the maximum slope predicted by the affine theory when using the same crosslink density as in the simulation (\cref{figkvssigmatheory}). There are, however, differences between theory and simulation in terms of slope profiles 
since various assumption are made by the theory that do not hold in the simulations: A randomly generated network does not have a uniform crosslink density along the filaments, these systems are prestressed, and there is no perfect isotropy. Moreover, the networks do not deform perfectly affinely.

\subsection{Nonaffinity}
\begin{figure}
  \centering
  \includegraphics{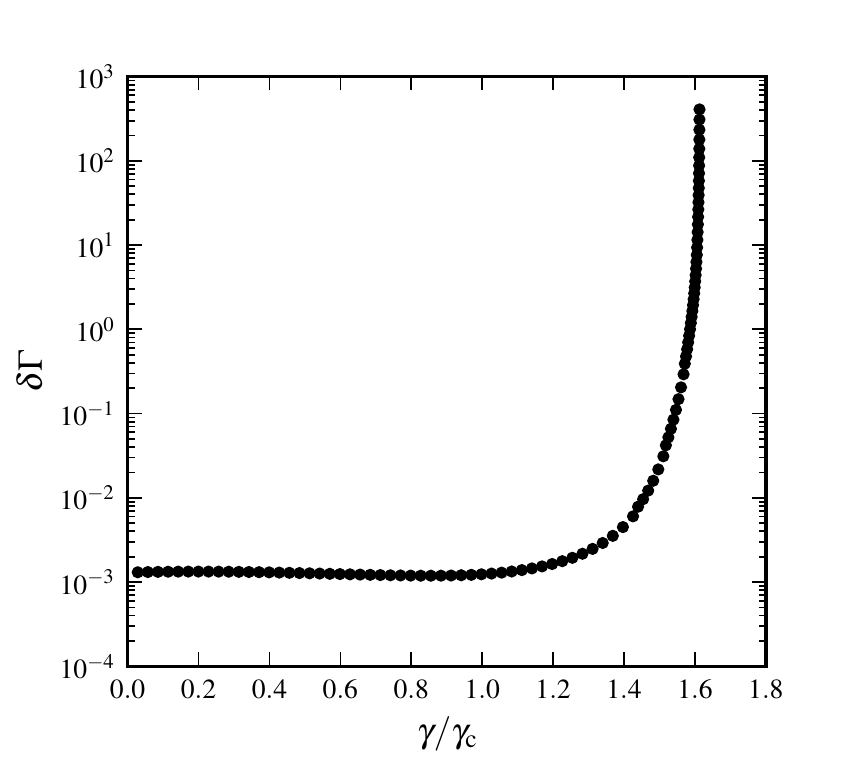}
  \caption{Differential nonaffinity $\delta \Gamma$ as a function of scaled shear strain $\gamma/\gamma_\text{c}$ for a system with $N=3000$, $n=60$, $L=0.3$, $l_0=0.06$, $\alpha=0.05$.}
  \label{fig:nonaffinity}
\end{figure}
In order to study to what extent simulation results deviate from affine theory, apart from prestress, nonuniform crosslink density, and anisotropy, we analyze the nonaffinity of the network deformation under shear.
For a single filament, we define its differential nonaffinity with respect to the center of mass by
\begin{align}
   \frac{\| \delta r_\text{aff} -\delta r \|^2}{\|\delta\gamma\|^2}\ ,
\end{align}
where $\delta r_\text{aff}$ and $\delta r$ are the 3D coordinates of a filament's center of mass after applying an incremental shear strain $\delta \gamma$ without and with relaxation, respectively.

We let $\delta \Gamma$ denote the average of the differential nonaffinities over all filaments. 
Affine approximations imply $\delta \Gamma = 0$. \cref{fig:nonaffinity} shows that center of mass deformations are mostly affine for small strains. However, the differential nonaffinity increases starting at a strain around $\gamma_\text{c}$ and eventually diverges as $\gamma \to \gamma_\text{d}$.
This can be understood, since the networks are strain stiffening, such that small incremental strain can induce large increase in the forces of individual crosslinks, thereby inducing large local rearrangements during energy minimization.

\begin{figure}
  \centering
  \includegraphics{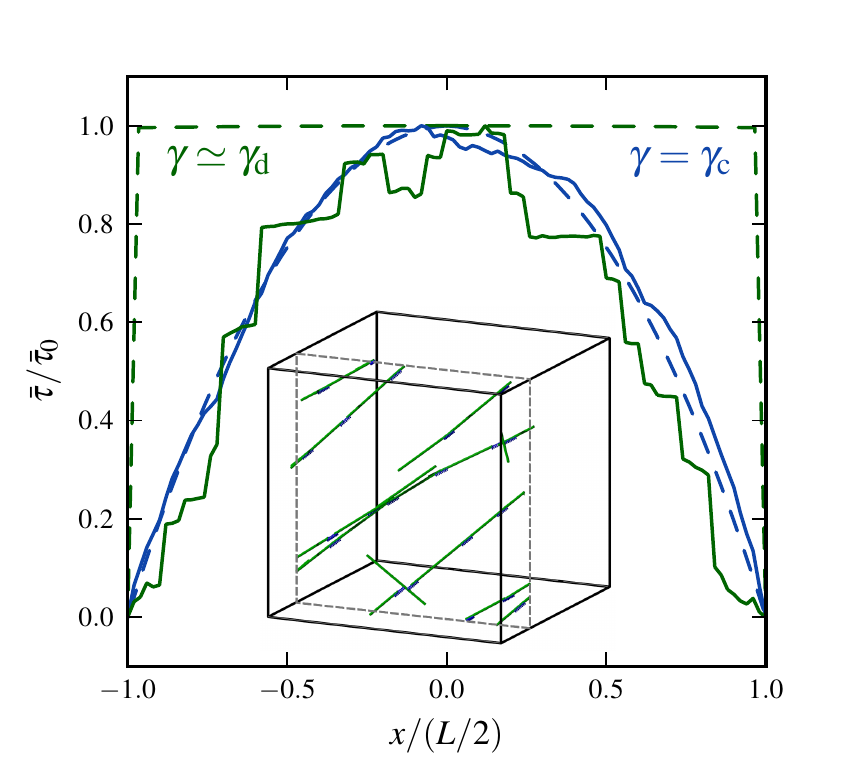}
  \caption{Average tension $\bar \tau$ as a function of position $x$ along the filament for various strain values. Tension $\bar\tau$ is normalized by its maximum absolute value $\bar\tau_0$. Dashed curves correspond to theoretical results for $n=60$ at $\gamma=\gamma_\text{c}$ (blue), $\gamma\simeq\gamma_\text{d}$ (green). Solid curves show simulation data, with $N=3000$, $n=60$, $L=0.3$, $l_0=0.06$, $\alpha=0.5$. Inset shows a snapshot of the same system at maximum strain $\gamma_\text{d}\simeq0.6$ where only the 15 most stretched crosslinks and the corresponding filaments are shown. They form singular paths that span the whole system, thereby preventing further stress reduction via nonaffine rearrangements in these finite systems.}
  \label{figTensionProfile}
\end{figure}

While increasing shear strain, there are \emph{force chains} \cite{Heussinger2007e,Huisman2007,Zagar2011} developing in the network, which carry most of the tension, and which cannot reduce their strain due to the fact that they span the entire system (see~inset of~\cref{figTensionProfile}).
We quantify this effect by considering tension profiles along filaments. The tension $\tau$ at position $x$ along a filament is given via $\tau(x)=\sum_{|x_i|>|x|} f_\text{cl}(u_i)$, where $\{x_i\}$ are the crosslink binding sites and $\{u_i\}$ their extensions ($u_i = \epsilon x_i$ in affine theory). \cref{figTensionProfile} shows tension profiles averaged over all filaments for both, theoretical and simulated systems at various strains.
In the simulations there is non-zero tension at zero strain due to prestress.
With increasing $\gamma$, the simulations resemble the profiles expected from affine theory. However, when approaching the maximum strain $\gamma_\text{d}$, the emergence of selective paths (force chains) that carry most of the tension becomes evident. The highly stretched crosslinks dominate the averaged tension profiles and therefore lead to jumps in the tension curves at the respective binding sites along the filament (green solid curve in~\cref{figTensionProfile}).
\subsection{Bending} 
\label{sub:Bending}
\begin{figure}
  \centering
  \includegraphics{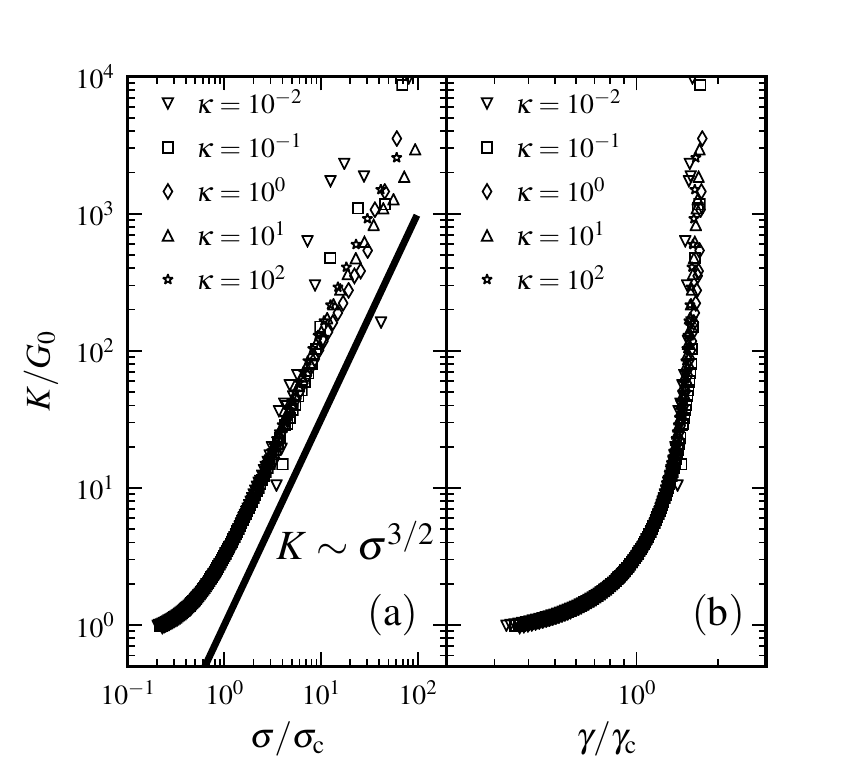}
  \caption{(a) Differential modulus $K$ as a function of shear stress $\sigma$, rescaled by linear modulus $G_0$ and critical stress $\sigma_c=\sigma(\gamma_c)$, respectively, for various bending rigidities $\kappa$. Solid straight line indicates power law scaling $K\sim \sigma^{3/2}$. 
  (b) Differential modulus $K$ as a function of shear strain $\gamma$, rescaled by linear modulus $G_0$ and critical strain $\gamma_\text{c}$, respectively. Parameters: $N=800$, $L=1$, $l_0=0.1$, system-size $L_x = L_y = 6$.}
  \label{bending}
\end{figure}

Thus far we have restricted our theory and simulations to rigid filaments that can neither bend nor stretch. 
In Ref.~\cite{Sharma2013a}, the authors considered finite stretching compliance of filaments, while bending compliance was assumed to be zero. They report that finite stretching stiffness does not impact the nonlinear stiffening regime of a composite network apart from the expected convergence (to some constant value) of the modulus at high strains.
Here we complement this analysis by considering filaments that have finite bending but no stretching compliance. We performed simulations on a 2D network because of the relative computational ease compared to the 3D case.
\begin{figure}
  \centering
  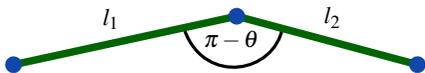
  \caption{Sketch of the local bending geometry of a filament (green) with crosslinks attached (blue). The local bending energy is given by $E_\text{b} =\kappa \theta^2/(2 l_\text{av})$, with $\kappa$ being the bending rigidity and $l_\text{av}=(l_1+l_2)/2$.}
  \label{figBendingAngle}
\end{figure}
In addition to the energy stored in the crosslinks, we consider bending energy of the form $E_\text{b} = \kappa \theta^2/ (2l_{\text{av}})$, where $\kappa$ is the bending rigidity, $\theta$ is the angle through which the filaments bend locally, and $l_{\text{av}}=(l_1+l_2)/2$ is the average distance between two adjacent pairs of crosslinks. We show the results in \cref{bending}.
The range of bending rigidity was chosen such that the linear modulus was still determined by the soft stretching modes of the crosslinks, so that 
bending did not impact the linear regime. As can be seen from these plots, bending compliance does not impact the nonlinear stiffening regime either---since
bending modes are \emph{geometrically} prohibited for large strains.

Thus, in isolation, neither bending nor stretching compliance of filaments impacts the nonlinear stiffening regime of composite networks.
These findings 
suggest that the theoretical models at present cannot explain the $K \sim \sigma$ scaling observed in experiments.
\section{Conclusions}
We have studied the elastic properties of composite crosslinked filamentous networks in 3D analytically and numerically. We modeled such networks as a collection of rigid filaments connected by WLC crosslinks.

Based on the affine theory introduced in Ref.~\cite{Sharma2013a} we derived asymptotic power law scaling exponents for the differential elastic modulus with stress, 
in the limit of infinite
crosslink density. In this case, the scaling exponents depend on the dimensionality of the system. In particular, we showed that 3D systems no longer exhibit a power law.
Furthermore, we showed that for \emph{finite} crosslink densities, the only persistent regime (over several orders of magnitude of stress) is the $\sigma^{3/2}$ scaling, as it is derived from the single WLC force-extension relation \cref{eq:wlcforce}.
This is in sharp contrast with the model proposed in Ref.~\cite{Broedersz2008,Broedersz2009b}, where linear scaling was suggested, independent of the dimensionality of the system. There model implies finite stress at \emph{any} strain and therefore does not apply to composite networks of rigid filaments with flexible crosslinks of finite length.

We further developed a simulation framework that allows us to measure the elastic response of random filamentous networks with WLC crosslinks. One important property of
these 3D networks is that, by construction, they are prestressed due to initial extensions of the crosslinks.
In addition to geometrical constraints, active elements such as motors can induce prestress as well \cite{Koenderink2009}.
We showed that the prestress in a network can dominate the linear response and might therefore be a feature that is worthwhile analyzing in experimental systems.

 Regarding nonlinear response, we observed divergence of stress (and differential modulus) at \emph{finite} strain. Close to this strain we measured a power law scaling of the differential modulus with stress, with an exponent 3/2, just as expected for a single WLC. In an intermediate-stress regime we observed local exponents that span the entire range of theoretically derived values for systems of differing dimensionality. The fact that our simulation results do not always resemble the predictions of a \emph{3D} affine theory, in this intermediate regime, may be attributed to nonaffine deformations.
Extracting the exact set of assumptions---such as uniform crosslink density, isotropy, or zero prestress---that are responsible for these discrepancies is left for future investigation.

Experiments (see, e.g.,~\cite{Gardel2006a,Kasza2009,Kasza2010}) have shown that in the nonlinear regime the differential modulus scales approximately linearly with the shear stress. We did not find such a regime in our simulations---neither when working with rigid filaments nor when incorporating finite bending stiffness (or enthalpic stretching as done in Ref.~\cite{Sharma2013a}). 
Therefore, we argue that none of the currently available theories can adequately explain the linear scaling of the differential modulus observed experimentally.
It could possibly be that the WLC model does not accurately describe the elastic response of a single crosslink throughout the whole experimentally accessible regime.
We speculate, however, that the linear scaling might be due to thermal fluctuations of the filaments, which have not been considered so far.

\section*{Acknowledgments}
The authors would like to thank Fred MacKintosh for fruitful discussions. This work was funded by the Deutsche For\-schungs\-ge\-mein\-schaft (DFG) within the collaborative research center SFB 755, project A3.
\appendix
\section{Derivation of scaling relations for the shear modulus}
\subsection{1D network}
\label{sub1Dappendix}
The integral \cref{eqEinfinite} for the total energy of a single filament can be solved to give
\begin{align}
  E_\text{1D}(\epsilon) = 2 \frac{n}{L} \left[ \frac{L^3\epsilon^2}{48l_0} - \frac{L^2\epsilon}{32} - \frac{l_0L}{8}-\frac{l_0^2}{4\epsilon}\ln\left(1-\frac{\epsilon L}{2l_0}\right)\right] \ .
\end{align}
The divergence of the energy for $\epsilon \to \epsilon_\text{d}=2 l_0/L$ stems from the term $\sim \frac{1}{\epsilon} \ln \left(1-\frac{\epsilon }{\epsilon_\text{d}}\right)$, which is therefore the only one that we need to consider for  the asymptotic scaling analysis in 2D and 3D.
\subsection{3D network} 
\label{sub:3D network}
To approximate the solution of the integral in \cref{eq:3Dstress}
we first carry out the $\phi$ integration analytically and obtain
\begin{align}
  \begin{split}
  \langle \sigma_\text{3D}\rangle_{\theta,\phi}(\gamma) &\sim
  \int_0^{\pi/2} \frac{\arctan\left[\sqrt{\frac{1+(\gamma/\gamma_\text{d}) \sin 2\theta}{1-(\gamma/\gamma_\text{d})\sin 2\theta}}\right] }{\sqrt{1-(\gamma/\gamma_\text{d})^2\sin^2 2\theta}}\\
  &\times \sin \theta\, d\theta \ .
\end{split}
\end{align}
The integral diverges for $\gamma=\gamma_\text{d}$ due to a pole at $\theta = \pi/4$. We can approximately consider $\tan^{-1}\left[\sqrt{\tfrac{1+(\gamma/\gamma_\text{d})\sin 2\theta}{1-(\gamma/\gamma_\text{d})\sin 2\theta}}\right] \times \sin \theta$ as a constant because it takes finite values around the pole.
Since we are interested in the regime close to the divergence of the integrand, we expand $\sin^2 2\theta$ up to second order in  $\nu\defeq \theta - \pi/4$.
We arrive at
\begin{align}
  \int_{-\pi/4}^{\pi/4} \frac{d\nu}{\sqrt{1-(\gamma/\gamma_\text{d})^2(1-4\nu^2)}} \ .
\end{align}
Approximation errors close to the boundary of the interval of integration that are made by expanding $\sin^2 2\theta$ are negligible, regarding the asymptotics, because the integrand diverges right at the center of the interval.
Now we define $\mu\defeq 1-\gamma/\gamma_\text{d}$ and drop all terms of higher than first order in $\mu$, since we are interested in the behavior close to $\gamma=\gamma_\text{d}$.
With $\eta^2\defeq 4\nu^2$ and $\delta\defeq 2\mu$, we obtain
\begin{align}
  \int_{-\pi/2}^{\pi/2} \frac{d\eta}{\sqrt{\eta^2(1-\delta)+\delta }} \ .
\end{align}
This can be integrated, with the diverging part being
\begin{align}
  &\sim\ln\left.\left(2\sqrt{\eta^2(1-\delta)^2+\delta(1-\delta)}+2(1-\delta)\eta\right)\right|^{\pi/2}_{-\pi/2} \ ,\\
  &\sim -\ln \delta,\\ &\sim -\ln(1-\gamma/\gamma_\text{d})\ ,
\end{align}
which is what has been proposed in \cref{sub:3D}.

\bibliography{library,additionalRefs}

\begin{thebibliography}{44}%
\makeatletter
\providecommand \@ifxundefined [1]{%
 \@ifx{#1\undefined}
}%
\providecommand \@ifnum [1]{%
 \ifnum #1\expandafter \@firstoftwo
 \else \expandafter \@secondoftwo
 \fi
}%
\providecommand \@ifx [1]{%
 \ifx #1\expandafter \@firstoftwo
 \else \expandafter \@secondoftwo
 \fi
}%
\providecommand \natexlab [1]{#1}%
\providecommand \enquote  [1]{``#1''}%
\providecommand \bibnamefont  [1]{#1}%
\providecommand \bibfnamefont [1]{#1}%
\providecommand \citenamefont [1]{#1}%
\providecommand \href@noop [0]{\@secondoftwo}%
\providecommand \href [0]{\begingroup \@sanitize@url \@href}%
\providecommand \@href[1]{\@@startlink{#1}\@@href}%
\providecommand \@@href[1]{\endgroup#1\@@endlink}%
\providecommand \@sanitize@url [0]{\catcode `\\12\catcode `\$12\catcode
  `\&12\catcode `\#12\catcode `\^12\catcode `\_12\catcode `\%12\relax}%
\providecommand \@@startlink[1]{}%
\providecommand \@@endlink[0]{}%
\providecommand \url  [0]{\begingroup\@sanitize@url \@url }%
\providecommand \@url [1]{\endgroup\@href {#1}{\urlprefix }}%
\providecommand \urlprefix  [0]{URL }%
\providecommand \Eprint [0]{\href }%
\providecommand \doibase [0]{http://dx.doi.org/}%
\providecommand \selectlanguage [0]{\@gobble}%
\providecommand \bibinfo  [0]{\@secondoftwo}%
\providecommand \bibfield  [0]{\@secondoftwo}%
\providecommand \translation [1]{[#1]}%
\providecommand \BibitemOpen [0]{}%
\providecommand \bibitemStop [0]{}%
\providecommand \bibitemNoStop [0]{.\EOS\space}%
\providecommand \EOS [0]{\spacefactor3000\relax}%
\providecommand \BibitemShut  [1]{\csname bibitem#1\endcsname}%
\let\auto@bib@innerbib\@empty
\bibitem [{\citenamefont {Janmey}\ \emph {et~al.}(1991)\citenamefont {Janmey},
  \citenamefont {Euteneuer}, \citenamefont {Traub},\ and\ \citenamefont
  {Schliwa}}]{Janmey1991}%
  \BibitemOpen
  \bibfield  {author} {\bibinfo {author} {\bibfnamefont {P.~A.}\ \bibnamefont
  {Janmey}}, \bibinfo {author} {\bibfnamefont {U.}~\bibnamefont {Euteneuer}},
  \bibinfo {author} {\bibfnamefont {P.}~\bibnamefont {Traub}}, \ and\ \bibinfo
  {author} {\bibfnamefont {M.}~\bibnamefont {Schliwa}},\ }\href
  {http://www.jcb.org/cgi/doi/10.1083/jcb.113.1.155
  http://www.pubmedcentral.nih.gov/articlerender.fcgi?artid=2288924\&tool=pmcentrez\&rendertype=abstract}
  {\bibfield  {journal} {\bibinfo  {journal} {J. Cell Biol.}\ }\textbf
  {\bibinfo {volume} {113}},\ \bibinfo {pages} {155} (\bibinfo {year}
  {1991})}\BibitemShut {NoStop}%
\bibitem [{\citenamefont {MacKintosh}\ and\ \citenamefont
  {Janmey}(1997)}]{MacKintosh1997}%
  \BibitemOpen
  \bibfield  {author} {\bibinfo {author} {\bibfnamefont {F.~C.}\ \bibnamefont
  {MacKintosh}}\ and\ \bibinfo {author} {\bibfnamefont {P.~A.}\ \bibnamefont
  {Janmey}},\ }\href {\doibase 10.1016/S1359-0286(97)80127-1} {\bibfield
  {journal} {\bibinfo  {journal} {Current Opinion in Solid State and Materials
  Science}\ }\textbf {\bibinfo {volume} {2}},\ \bibinfo {pages} {350} (\bibinfo
  {year} {1997})}\BibitemShut {NoStop}%
\bibitem [{\citenamefont {Xu}\ \emph {et~al.}(1998)\citenamefont {Xu},
  \citenamefont {Wirtz},\ and\ \citenamefont {Pollard}}]{Xu1998}%
  \BibitemOpen
  \bibfield  {author} {\bibinfo {author} {\bibfnamefont {J.}~\bibnamefont
  {Xu}}, \bibinfo {author} {\bibfnamefont {D.}~\bibnamefont {Wirtz}}, \ and\
  \bibinfo {author} {\bibfnamefont {T.~D.}\ \bibnamefont {Pollard}},\ }\href
  {\doibase 10.1074/jbc.273.16.9570} {\bibfield  {journal} {\bibinfo  {journal}
  {J. Biol. Chem.}\ }\textbf {\bibinfo {volume} {273}},\ \bibinfo {pages}
  {9570} (\bibinfo {year} {1998})}\BibitemShut {NoStop}%
\bibitem [{\citenamefont {Gardel}\ \emph
  {et~al.}(2004{\natexlab{a}})\citenamefont {Gardel}, \citenamefont {Shin},
  \citenamefont {MacKintosh}, \citenamefont {Mahadevan}, \citenamefont
  {Matsudaira},\ and\ \citenamefont {Weitz}}]{Gardel2004}%
  \BibitemOpen
  \bibfield  {author} {\bibinfo {author} {\bibfnamefont {M.~L.}\ \bibnamefont
  {Gardel}}, \bibinfo {author} {\bibfnamefont {J.~H.}\ \bibnamefont {Shin}},
  \bibinfo {author} {\bibfnamefont {F.~C.}\ \bibnamefont {MacKintosh}},
  \bibinfo {author} {\bibfnamefont {L.}~\bibnamefont {Mahadevan}}, \bibinfo
  {author} {\bibfnamefont {P.}~\bibnamefont {Matsudaira}}, \ and\ \bibinfo
  {author} {\bibfnamefont {D.~A.}\ \bibnamefont {Weitz}},\ }\href {\doibase
  10.1126/science.1095087} {\bibfield  {journal} {\bibinfo  {journal}
  {Science}\ }\textbf {\bibinfo {volume} {304}},\ \bibinfo {pages} {1301}
  (\bibinfo {year} {2004}{\natexlab{a}})}\BibitemShut {NoStop}%
\bibitem [{\citenamefont {Gardel}\ \emph
  {et~al.}(2004{\natexlab{b}})\citenamefont {Gardel}, \citenamefont {Shin},
  \citenamefont {MacKintosh}, \citenamefont {Mahadevan}, \citenamefont
  {Matsudaira},\ and\ \citenamefont {Weitz}}]{Gardel2004b}%
  \BibitemOpen
  \bibfield  {author} {\bibinfo {author} {\bibfnamefont {M.~L.}\ \bibnamefont
  {Gardel}}, \bibinfo {author} {\bibfnamefont {J.~H.}\ \bibnamefont {Shin}},
  \bibinfo {author} {\bibfnamefont {F.~C.}\ \bibnamefont {MacKintosh}},
  \bibinfo {author} {\bibfnamefont {L.}~\bibnamefont {Mahadevan}}, \bibinfo
  {author} {\bibfnamefont {P.~A.}\ \bibnamefont {Matsudaira}}, \ and\ \bibinfo
  {author} {\bibfnamefont {D.~A.}\ \bibnamefont {Weitz}},\ }\href {\doibase
  10.1103/PhysRevLett.93.188102} {\bibfield  {journal} {\bibinfo  {journal}
  {Phys. Rev. Lett.}\ }\textbf {\bibinfo {volume} {93}},\ \bibinfo {pages} {1}
  (\bibinfo {year} {2004}{\natexlab{b}})}\BibitemShut {NoStop}%
\bibitem [{\citenamefont {Storm}\ \emph {et~al.}(2005)\citenamefont {Storm},
  \citenamefont {Pastore}, \citenamefont {Mackintosh}, \citenamefont
  {Lubensky},\ and\ \citenamefont {Janmey}}]{Storm2005}%
  \BibitemOpen
  \bibfield  {author} {\bibinfo {author} {\bibfnamefont {C.}~\bibnamefont
  {Storm}}, \bibinfo {author} {\bibfnamefont {J.~J.}\ \bibnamefont {Pastore}},
  \bibinfo {author} {\bibfnamefont {F.~C.}\ \bibnamefont {Mackintosh}},
  \bibinfo {author} {\bibfnamefont {T.~C.}\ \bibnamefont {Lubensky}}, \ and\
  \bibinfo {author} {\bibfnamefont {P.~A.}\ \bibnamefont {Janmey}},\ }\href
  {\doibase 10.1038/nature03497.1.} {\bibfield  {journal} {\bibinfo  {journal}
  {Nature}\ }\textbf {\bibinfo {volume} {435}},\ \bibinfo {pages} {191}
  (\bibinfo {year} {2005})}\BibitemShut {NoStop}%
\bibitem [{\citenamefont {Bausch}\ and\ \citenamefont
  {Kroy}(2006)}]{Bausch2006b}%
  \BibitemOpen
  \bibfield  {author} {\bibinfo {author} {\bibfnamefont {A.~R.}\ \bibnamefont
  {Bausch}}\ and\ \bibinfo {author} {\bibfnamefont {K.}~\bibnamefont {Kroy}},\
  }\href {\doibase 10.1038/nphys260} {\bibfield  {journal} {\bibinfo  {journal}
  {Nature Physics}\ }\textbf {\bibinfo {volume} {2}},\ \bibinfo {pages} {231}
  (\bibinfo {year} {2006})}\BibitemShut {NoStop}%
\bibitem [{\citenamefont {Koenderink}\ \emph {et~al.}(2006)\citenamefont
  {Koenderink}, \citenamefont {Atakhorrami}, \citenamefont {MacKintosh},\ and\
  \citenamefont {Schmidt}}]{Koenderink2006}%
  \BibitemOpen
  \bibfield  {author} {\bibinfo {author} {\bibfnamefont {G.~H.}\ \bibnamefont
  {Koenderink}}, \bibinfo {author} {\bibfnamefont {M.}~\bibnamefont
  {Atakhorrami}}, \bibinfo {author} {\bibfnamefont {F.~C.}\ \bibnamefont
  {MacKintosh}}, \ and\ \bibinfo {author} {\bibfnamefont {C.~F.}\ \bibnamefont
  {Schmidt}},\ }\href {\doibase 10.1103/PhysRevLett.96.138307} {\bibfield
  {journal} {\bibinfo  {journal} {Phys. Rev. Lett.}\ }\textbf {\bibinfo
  {volume} {96}},\ \bibinfo {pages} {138307} (\bibinfo {year}
  {2006})}\BibitemShut {NoStop}%
\bibitem [{\citenamefont {Chaudhuri}\ \emph {et~al.}(2007)\citenamefont
  {Chaudhuri}, \citenamefont {Parekh},\ and\ \citenamefont
  {Fletcher}}]{Chaudhuri2007}%
  \BibitemOpen
  \bibfield  {author} {\bibinfo {author} {\bibfnamefont {O.}~\bibnamefont
  {Chaudhuri}}, \bibinfo {author} {\bibfnamefont {S.~H.}\ \bibnamefont
  {Parekh}}, \ and\ \bibinfo {author} {\bibfnamefont {D.~A.}\ \bibnamefont
  {Fletcher}},\ }\href {\doibase 10.1038/nature05459} {\bibfield  {journal}
  {\bibinfo  {journal} {Nature}\ }\textbf {\bibinfo {volume} {445}},\ \bibinfo
  {pages} {295} (\bibinfo {year} {2007})}\BibitemShut {NoStop}%
\bibitem [{\citenamefont {Janmey}\ \emph {et~al.}(2007)\citenamefont {Janmey},
  \citenamefont {McCormick}, \citenamefont {Rammensee}, \citenamefont {Leight},
  \citenamefont {Georges},\ and\ \citenamefont {MacKintosh}}]{Janmey2007a}%
  \BibitemOpen
  \bibfield  {author} {\bibinfo {author} {\bibfnamefont {P.~A.}\ \bibnamefont
  {Janmey}}, \bibinfo {author} {\bibfnamefont {M.~E.}\ \bibnamefont
  {McCormick}}, \bibinfo {author} {\bibfnamefont {S.}~\bibnamefont
  {Rammensee}}, \bibinfo {author} {\bibfnamefont {J.~L.}\ \bibnamefont
  {Leight}}, \bibinfo {author} {\bibfnamefont {P.~C.}\ \bibnamefont {Georges}},
  \ and\ \bibinfo {author} {\bibfnamefont {F.~C.}\ \bibnamefont {MacKintosh}},\
  }\href {\doibase 10.1038/nmat1810} {\bibfield  {journal} {\bibinfo  {journal}
  {Nature Materials}\ }\textbf {\bibinfo {volume} {6}},\ \bibinfo {pages} {48}
  (\bibinfo {year} {2007})}\BibitemShut {NoStop}%
\bibitem [{\citenamefont {Kasza}\ \emph {et~al.}(2007)\citenamefont {Kasza},
  \citenamefont {Rowat}, \citenamefont {Liu}, \citenamefont {Angelini},
  \citenamefont {Brangwynne}, \citenamefont {Koenderink},\ and\ \citenamefont
  {Weitz}}]{Kasza2007}%
  \BibitemOpen
  \bibfield  {author} {\bibinfo {author} {\bibfnamefont {K.~E.}\ \bibnamefont
  {Kasza}}, \bibinfo {author} {\bibfnamefont {A.~C.}\ \bibnamefont {Rowat}},
  \bibinfo {author} {\bibfnamefont {J.}~\bibnamefont {Liu}}, \bibinfo {author}
  {\bibfnamefont {T.~E.}\ \bibnamefont {Angelini}}, \bibinfo {author}
  {\bibfnamefont {C.~P.}\ \bibnamefont {Brangwynne}}, \bibinfo {author}
  {\bibfnamefont {G.~H.}\ \bibnamefont {Koenderink}}, \ and\ \bibinfo {author}
  {\bibfnamefont {D.~A.}\ \bibnamefont {Weitz}},\ }\href {\doibase
  10.1016/j.ceb.2006.12.002} {\bibfield  {journal} {\bibinfo  {journal} {Curr.
  Opin. Cell Biol.}\ }\textbf {\bibinfo {volume} {19}},\ \bibinfo {pages} {101}
  (\bibinfo {year} {2007})}\BibitemShut {NoStop}%
\bibitem [{\citenamefont {Liu}\ \emph {et~al.}(2007)\citenamefont {Liu},
  \citenamefont {Koenderink}, \citenamefont {Kasza}, \citenamefont
  {MacKintosh},\ and\ \citenamefont {Weitz}}]{Liu2007}%
  \BibitemOpen
  \bibfield  {author} {\bibinfo {author} {\bibfnamefont {J.}~\bibnamefont
  {Liu}}, \bibinfo {author} {\bibfnamefont {G.~H.}\ \bibnamefont {Koenderink}},
  \bibinfo {author} {\bibfnamefont {K.~E.}\ \bibnamefont {Kasza}}, \bibinfo
  {author} {\bibfnamefont {F.~C.}\ \bibnamefont {MacKintosh}}, \ and\ \bibinfo
  {author} {\bibfnamefont {D.~A.}\ \bibnamefont {Weitz}},\ }\href {\doibase
  10.1103/PhysRevLett.98.198304} {\bibfield  {journal} {\bibinfo  {journal}
  {Phys. Rev. Lett.}\ }\textbf {\bibinfo {volume} {98}},\ \bibinfo {pages}
  {198304} (\bibinfo {year} {2007})}\BibitemShut {NoStop}%
\bibitem [{\citenamefont {Gardel}\ \emph
  {et~al.}(2006{\natexlab{a}})\citenamefont {Gardel}, \citenamefont {Nakamura},
  \citenamefont {Hartwig}, \citenamefont {Crocker}, \citenamefont {Stossel},\
  and\ \citenamefont {Weitz}}]{Gardel2006a}%
  \BibitemOpen
  \bibfield  {author} {\bibinfo {author} {\bibfnamefont {M.~L.}\ \bibnamefont
  {Gardel}}, \bibinfo {author} {\bibfnamefont {F.}~\bibnamefont {Nakamura}},
  \bibinfo {author} {\bibfnamefont {J.~H.}\ \bibnamefont {Hartwig}}, \bibinfo
  {author} {\bibfnamefont {J.~C.}\ \bibnamefont {Crocker}}, \bibinfo {author}
  {\bibfnamefont {T.~P.}\ \bibnamefont {Stossel}}, \ and\ \bibinfo {author}
  {\bibfnamefont {D.~A.}\ \bibnamefont {Weitz}},\ }\href {\doibase
  10.1073/pnas.0504777103} {\bibfield  {journal} {\bibinfo  {journal} {Proc.
  Natl. Acad. Sci. U.S.A.}\ }\textbf {\bibinfo {volume} {103}},\ \bibinfo
  {pages} {1762} (\bibinfo {year} {2006}{\natexlab{a}})}\BibitemShut {NoStop}%
\bibitem [{\citenamefont {Gardel}\ \emph
  {et~al.}(2006{\natexlab{b}})\citenamefont {Gardel}, \citenamefont {Nakamura},
  \citenamefont {Hartwig}, \citenamefont {Crocker}, \citenamefont {Stossel},\
  and\ \citenamefont {Weitz}}]{Gardel2006}%
  \BibitemOpen
  \bibfield  {author} {\bibinfo {author} {\bibfnamefont {M.~L.}\ \bibnamefont
  {Gardel}}, \bibinfo {author} {\bibfnamefont {F.}~\bibnamefont {Nakamura}},
  \bibinfo {author} {\bibfnamefont {J.}~\bibnamefont {Hartwig}}, \bibinfo
  {author} {\bibfnamefont {J.~C.}\ \bibnamefont {Crocker}}, \bibinfo {author}
  {\bibfnamefont {T.~P.}\ \bibnamefont {Stossel}}, \ and\ \bibinfo {author}
  {\bibfnamefont {D.~A.}\ \bibnamefont {Weitz}},\ }\href
  {http://www.ncbi.nlm.nih.gov/pubmed/16606229} {\bibfield  {journal} {\bibinfo
   {journal} {Phys. Rev. Lett.}\ }\textbf {\bibinfo {volume} {96}},\ \bibinfo
  {pages} {088102} (\bibinfo {year} {2006}{\natexlab{b}})}\BibitemShut
  {NoStop}%
\bibitem [{\citenamefont {DiDonna}\ and\ \citenamefont
  {Levine}(2007)}]{DiDonna2007}%
  \BibitemOpen
  \bibfield  {author} {\bibinfo {author} {\bibfnamefont {B.~A.}\ \bibnamefont
  {DiDonna}}\ and\ \bibinfo {author} {\bibfnamefont {A.~J.}\ \bibnamefont
  {Levine}},\ }\href {\doibase 10.1103/PhysRevE.75.041909} {\bibfield
  {journal} {\bibinfo  {journal} {Phys. Rev. E: Stat., Nonlinear, Soft Matter
  Phys.}\ }\textbf {\bibinfo {volume} {75}},\ \bibinfo {pages} {041909}
  (\bibinfo {year} {2007})}\BibitemShut {NoStop}%
\bibitem [{\citenamefont {Broedersz}\ \emph {et~al.}(2008)\citenamefont
  {Broedersz}, \citenamefont {Storm},\ and\ \citenamefont
  {MacKintosh}}]{Broedersz2008}%
  \BibitemOpen
  \bibfield  {author} {\bibinfo {author} {\bibfnamefont {C.~P.}\ \bibnamefont
  {Broedersz}}, \bibinfo {author} {\bibfnamefont {C.}~\bibnamefont {Storm}}, \
  and\ \bibinfo {author} {\bibfnamefont {F.~C.}\ \bibnamefont {MacKintosh}},\
  }\href {\doibase 10.1103/PhysRevLett.101.118103} {\bibfield  {journal}
  {\bibinfo  {journal} {Phys. Rev. Lett.}\ }\textbf {\bibinfo {volume} {101}},\
  \bibinfo {pages} {118103} (\bibinfo {year} {2008})}\BibitemShut {NoStop}%
\bibitem [{\citenamefont {Dalhaimer}\ \emph {et~al.}(2007)\citenamefont
  {Dalhaimer}, \citenamefont {Discher},\ and\ \citenamefont
  {Lubensky}}]{Dalhaimer2007}%
  \BibitemOpen
  \bibfield  {author} {\bibinfo {author} {\bibfnamefont {P.}~\bibnamefont
  {Dalhaimer}}, \bibinfo {author} {\bibfnamefont {D.~E.}\ \bibnamefont
  {Discher}}, \ and\ \bibinfo {author} {\bibfnamefont {T.~C.}\ \bibnamefont
  {Lubensky}},\ }\href {\doibase 10.1038/nphys567} {\bibfield  {journal}
  {\bibinfo  {journal} {Nature Physics}\ }\textbf {\bibinfo {volume} {3}},\
  \bibinfo {pages} {354} (\bibinfo {year} {2007})}\BibitemShut {NoStop}%
\bibitem [{\citenamefont {Lee}\ \emph {et~al.}(2009)\citenamefont {Lee},
  \citenamefont {Pelz}, \citenamefont {Ferrer}, \citenamefont {Kim},
  \citenamefont {Lang},\ and\ \citenamefont {Kamm}}]{Lee2009}%
  \BibitemOpen
  \bibfield  {author} {\bibinfo {author} {\bibfnamefont {H.}~\bibnamefont
  {Lee}}, \bibinfo {author} {\bibfnamefont {B.}~\bibnamefont {Pelz}}, \bibinfo
  {author} {\bibfnamefont {J.~M.}\ \bibnamefont {Ferrer}}, \bibinfo {author}
  {\bibfnamefont {T.}~\bibnamefont {Kim}}, \bibinfo {author} {\bibfnamefont
  {M.~J.}\ \bibnamefont {Lang}}, \ and\ \bibinfo {author} {\bibfnamefont
  {R.~D.}\ \bibnamefont {Kamm}},\ }\href@noop {} {\bibfield  {journal}
  {\bibinfo  {journal} {Cellular and Molecular Bioengineering}\ }\textbf
  {\bibinfo {volume} {2}},\ \bibinfo {pages} {28} (\bibinfo {year}
  {2009})}\BibitemShut {NoStop}%
\bibitem [{\citenamefont {Sharma}\ \emph {et~al.}(2013)\citenamefont {Sharma},
  \citenamefont {Sheinman}, \citenamefont {Heidemann},\ and\ \citenamefont
  {MacKintosh}}]{Sharma2013a}%
  \BibitemOpen
  \bibfield  {author} {\bibinfo {author} {\bibfnamefont {A.}~\bibnamefont
  {Sharma}}, \bibinfo {author} {\bibfnamefont {M.}~\bibnamefont {Sheinman}},
  \bibinfo {author} {\bibfnamefont {K.~M.}\ \bibnamefont {Heidemann}}, \ and\
  \bibinfo {author} {\bibfnamefont {F.~C.}\ \bibnamefont {MacKintosh}},\ }\href
  {\doibase 10.1103/PhysRevE.88.052705} {\bibfield  {journal} {\bibinfo
  {journal} {Phys. Rev. E: Stat., Nonlinear, Soft Matter Phys.}\ }\textbf
  {\bibinfo {volume} {88}},\ \bibinfo {pages} {052705} (\bibinfo {year}
  {2013})}\BibitemShut {NoStop}%
\bibitem [{\citenamefont {Kasza}\ \emph {et~al.}(2009)\citenamefont {Kasza},
  \citenamefont {Koenderink}, \citenamefont {Lin}, \citenamefont {Broedersz},
  \citenamefont {Messner}, \citenamefont {Nakamura}, \citenamefont {Stossel},
  \citenamefont {MacKintosh},\ and\ \citenamefont {Weitz}}]{Kasza2009}%
  \BibitemOpen
  \bibfield  {author} {\bibinfo {author} {\bibfnamefont {K.~E.}\ \bibnamefont
  {Kasza}}, \bibinfo {author} {\bibfnamefont {G.~H.}\ \bibnamefont
  {Koenderink}}, \bibinfo {author} {\bibfnamefont {Y.~C.}\ \bibnamefont {Lin}},
  \bibinfo {author} {\bibfnamefont {C.~P.}\ \bibnamefont {Broedersz}}, \bibinfo
  {author} {\bibfnamefont {W.}~\bibnamefont {Messner}}, \bibinfo {author}
  {\bibfnamefont {F.}~\bibnamefont {Nakamura}}, \bibinfo {author}
  {\bibfnamefont {T.~P.}\ \bibnamefont {Stossel}}, \bibinfo {author}
  {\bibfnamefont {F.~C.}\ \bibnamefont {MacKintosh}}, \ and\ \bibinfo {author}
  {\bibfnamefont {D.~A.}\ \bibnamefont {Weitz}},\ }\href {\doibase
  10.1103/PhysRevE.79.041928} {\bibfield  {journal} {\bibinfo  {journal} {Phys.
  Rev. E: Stat., Nonlinear, Soft Matter Phys.}\ }\textbf {\bibinfo {volume}
  {79}},\ \bibinfo {pages} {041928} (\bibinfo {year} {2009})}\BibitemShut
  {NoStop}%
\bibitem [{\citenamefont {Kasza}\ \emph {et~al.}(2010)\citenamefont {Kasza},
  \citenamefont {Broedersz}, \citenamefont {Koenderink}, \citenamefont {Lin},
  \citenamefont {Messner}, \citenamefont {Millman}, \citenamefont {Nakamura},
  \citenamefont {Stossel}, \citenamefont {Mackintosh},\ and\ \citenamefont
  {Weitz}}]{Kasza2010}%
  \BibitemOpen
  \bibfield  {author} {\bibinfo {author} {\bibfnamefont {K.~E.}\ \bibnamefont
  {Kasza}}, \bibinfo {author} {\bibfnamefont {C.~P.}\ \bibnamefont
  {Broedersz}}, \bibinfo {author} {\bibfnamefont {G.~H.}\ \bibnamefont
  {Koenderink}}, \bibinfo {author} {\bibfnamefont {Y.~C.}\ \bibnamefont {Lin}},
  \bibinfo {author} {\bibfnamefont {W.}~\bibnamefont {Messner}}, \bibinfo
  {author} {\bibfnamefont {E.~A.}\ \bibnamefont {Millman}}, \bibinfo {author}
  {\bibfnamefont {F.}~\bibnamefont {Nakamura}}, \bibinfo {author}
  {\bibfnamefont {T.~P.}\ \bibnamefont {Stossel}}, \bibinfo {author}
  {\bibfnamefont {F.~C.}\ \bibnamefont {Mackintosh}}, \ and\ \bibinfo {author}
  {\bibfnamefont {D.~A.}\ \bibnamefont {Weitz}},\ }\href {\doibase
  10.1016/j.bpj.2010.06.025} {\bibfield  {journal} {\bibinfo  {journal}
  {Biophys. J.}\ }\textbf {\bibinfo {volume} {99}},\ \bibinfo {pages} {1091}
  (\bibinfo {year} {2010})}\BibitemShut {NoStop}%
\bibitem [{\citenamefont {Bustamante}\ \emph {et~al.}(1994)\citenamefont
  {Bustamante}, \citenamefont {Marko}, \citenamefont {Siggia},\ and\
  \citenamefont {Smith}}]{Bustamante1994}%
  \BibitemOpen
  \bibfield  {author} {\bibinfo {author} {\bibfnamefont {C.}~\bibnamefont
  {Bustamante}}, \bibinfo {author} {\bibfnamefont {J.~F.}\ \bibnamefont
  {Marko}}, \bibinfo {author} {\bibfnamefont {E.~D.}\ \bibnamefont {Siggia}}, \
  and\ \bibinfo {author} {\bibfnamefont {S.}~\bibnamefont {Smith}},\ }\href
  {\doibase 10.1126/science.8079175} {\bibfield  {journal} {\bibinfo  {journal}
  {Science}\ }\textbf {\bibinfo {volume} {265}},\ \bibinfo {pages} {1599}
  (\bibinfo {year} {1994})}\BibitemShut {NoStop}%
\bibitem [{\citenamefont {Marko}\ and\ \citenamefont
  {Siggia}(1995)}]{Marko1995a}%
  \BibitemOpen
  \bibfield  {author} {\bibinfo {author} {\bibfnamefont {J.~F.}\ \bibnamefont
  {Marko}}\ and\ \bibinfo {author} {\bibfnamefont {E.~D.}\ \bibnamefont
  {Siggia}},\ }\href {\doibase 10.1021/ma00130a008} {\bibfield  {journal}
  {\bibinfo  {journal} {Macromolecules}\ }\textbf {\bibinfo {volume} {28}},\
  \bibinfo {pages} {8759} (\bibinfo {year} {1995})}\BibitemShut {NoStop}%
\bibitem [{\citenamefont {Schwaiger}\ \emph {et~al.}(2004)\citenamefont
  {Schwaiger}, \citenamefont {Kardinal}, \citenamefont {Schleicher},
  \citenamefont {Noegel},\ and\ \citenamefont {Rief}}]{Schwaiger2004}%
  \BibitemOpen
  \bibfield  {author} {\bibinfo {author} {\bibfnamefont {I.}~\bibnamefont
  {Schwaiger}}, \bibinfo {author} {\bibfnamefont {A.}~\bibnamefont {Kardinal}},
  \bibinfo {author} {\bibfnamefont {M.}~\bibnamefont {Schleicher}}, \bibinfo
  {author} {\bibfnamefont {A.~A.}\ \bibnamefont {Noegel}}, \ and\ \bibinfo
  {author} {\bibfnamefont {M.}~\bibnamefont {Rief}},\ }\href {\doibase
  10.1038/nsmb705} {\bibfield  {journal} {\bibinfo  {journal} {Nature
  structural \& molecular biology}\ }\textbf {\bibinfo {volume} {11}},\
  \bibinfo {pages} {81} (\bibinfo {year} {2004})}\BibitemShut {NoStop}%
\bibitem [{\citenamefont {Furuike}\ \emph {et~al.}(2001)\citenamefont
  {Furuike}, \citenamefont {Ito},\ and\ \citenamefont
  {Yamazaki}}]{Furuike2001}%
  \BibitemOpen
  \bibfield  {author} {\bibinfo {author} {\bibfnamefont {S.}~\bibnamefont
  {Furuike}}, \bibinfo {author} {\bibfnamefont {T.}~\bibnamefont {Ito}}, \ and\
  \bibinfo {author} {\bibfnamefont {M.}~\bibnamefont {Yamazaki}},\ }\href
  {http://www.sciencedirect.com/science/article/pii/S0014579301024978}
  {\bibfield  {journal} {\bibinfo  {journal} {FEBS letters}\ }\textbf {\bibinfo
  {volume} {498}},\ \bibinfo {pages} {72} (\bibinfo {year} {2001})}\BibitemShut
  {NoStop}%
\bibitem [{\citenamefont {Broedersz}\ \emph {et~al.}(2009)\citenamefont
  {Broedersz}, \citenamefont {Storm},\ and\ \citenamefont
  {MacKintosh}}]{Broedersz2009b}%
  \BibitemOpen
  \bibfield  {author} {\bibinfo {author} {\bibfnamefont {C.~P.}\ \bibnamefont
  {Broedersz}}, \bibinfo {author} {\bibfnamefont {C.}~\bibnamefont {Storm}}, \
  and\ \bibinfo {author} {\bibfnamefont {F.~C.}\ \bibnamefont {MacKintosh}},\
  }\href {\doibase 10.1103/PhysRevE.79.061914} {\bibfield  {journal} {\bibinfo
  {journal} {Phys. Rev. E: Stat., Nonlinear, Soft Matter Phys.}\ }\textbf
  {\bibinfo {volume} {79}},\ \bibinfo {pages} {61914} (\bibinfo {year}
  {2009})}\BibitemShut {NoStop}%
\bibitem [{Note2()}]{Note2}%
  \BibitemOpen
  \bibinfo {note} {More precisely, it is a \protect \emph {free} energy, which
  includes both, energetic (bending) and entropic terms for the crosslinks (not
  for the filaments).}\BibitemShut {Stop}%
\bibitem [{\citenamefont {Landau}\ and\ \citenamefont
  {Lifshitz}(1975)}]{Landau1975elasticity}%
  \BibitemOpen
  \bibfield  {author} {\bibinfo {author} {\bibfnamefont {L.}~\bibnamefont
  {Landau}}\ and\ \bibinfo {author} {\bibfnamefont {E.}~\bibnamefont
  {Lifshitz}},\ }\href@noop {} {\emph {\bibinfo {title} {Elasticity theory}}}\
  (\bibinfo  {publisher} {Pergamon Press},\ \bibinfo {year} {1975})\BibitemShut
  {NoStop}%
\bibitem [{Note3()}]{Note3}%
  \BibitemOpen
  \bibinfo {note} {We do neither take into account fluctuations of the
  filaments nor excluded-volume effects.}\BibitemShut {Stop}%
\bibitem [{\citenamefont {W\"{a}chter}\ and\ \citenamefont
  {Biegler}(2005)}]{Wachter2006}%
  \BibitemOpen
  \bibfield  {author} {\bibinfo {author} {\bibfnamefont {A.}~\bibnamefont
  {W\"{a}chter}}\ and\ \bibinfo {author} {\bibfnamefont {L.~T.}\ \bibnamefont
  {Biegler}},\ }\href {\doibase 10.1007/s10107-004-0559-y} {\bibfield
  {journal} {\bibinfo  {journal} {Math. Prog.}\ }\textbf {\bibinfo {volume}
  {106}},\ \bibinfo {pages} {25} (\bibinfo {year} {2005})}\BibitemShut
  {NoStop}%
\bibitem [{\citenamefont {Lees}\ and\ \citenamefont
  {Edwards}(1972)}]{Lees1972}%
  \BibitemOpen
  \bibfield  {author} {\bibinfo {author} {\bibfnamefont {A.~W.}\ \bibnamefont
  {Lees}}\ and\ \bibinfo {author} {\bibfnamefont {S.~F.}\ \bibnamefont
  {Edwards}},\ }\href {\doibase 10.1088/0022-3719/5/15/006} {\bibfield
  {journal} {\bibinfo  {journal} {J. Phys. C: Solid State Phys.}\ }\textbf
  {\bibinfo {volume} {5}},\ \bibinfo {pages} {1921} (\bibinfo {year}
  {1972})}\BibitemShut {NoStop}%
\bibitem [{\citenamefont {Wilhelm}\ and\ \citenamefont
  {Frey}(2003)}]{Wilhelm2003d}%
  \BibitemOpen
  \bibfield  {author} {\bibinfo {author} {\bibfnamefont {J.}~\bibnamefont
  {Wilhelm}}\ and\ \bibinfo {author} {\bibfnamefont {E.}~\bibnamefont {Frey}},\
  }\href {\doibase 10.1103/PhysRevLett.91.108103} {\bibfield  {journal}
  {\bibinfo  {journal} {Phys. Rev. Lett.}\ }\textbf {\bibinfo {volume} {91}},\
  \bibinfo {pages} {108103} (\bibinfo {year} {2003})},\ \Eprint
  {http://arxiv.org/abs/0303592} {arXiv:0303592 [cond-mat]} \BibitemShut
  {NoStop}%
\bibitem [{\citenamefont {Head}\ \emph {et~al.}(2003)\citenamefont {Head},
  \citenamefont {Levine},\ and\ \citenamefont {MacKintosh}}]{Head2003}%
  \BibitemOpen
  \bibfield  {author} {\bibinfo {author} {\bibfnamefont {D.~A.}\ \bibnamefont
  {Head}}, \bibinfo {author} {\bibfnamefont {A.~J.}\ \bibnamefont {Levine}}, \
  and\ \bibinfo {author} {\bibfnamefont {F.~C.}\ \bibnamefont {MacKintosh}},\
  }\href {\doibase 10.1103/PhysRevLett.91.108102} {\bibfield  {journal}
  {\bibinfo  {journal} {Phys. Rev. Lett.}\ }\textbf {\bibinfo {volume} {91}},\
  \bibinfo {pages} {2} (\bibinfo {year} {2003})}\BibitemShut {NoStop}%
\bibitem [{\citenamefont {Onck}\ \emph {et~al.}(2005)\citenamefont {Onck},
  \citenamefont {Koeman}, \citenamefont {van Dillen},\ and\ \citenamefont
  {van~der Giessen}}]{Onck2005}%
  \BibitemOpen
  \bibfield  {author} {\bibinfo {author} {\bibfnamefont {P.~R.}\ \bibnamefont
  {Onck}}, \bibinfo {author} {\bibfnamefont {T.}~\bibnamefont {Koeman}},
  \bibinfo {author} {\bibfnamefont {T.}~\bibnamefont {van Dillen}}, \ and\
  \bibinfo {author} {\bibfnamefont {E.}~\bibnamefont {van~der Giessen}},\
  }\href {\doibase 10.1103/PhysRevLett.95.178102} {\bibfield  {journal}
  {\bibinfo  {journal} {Phys. Rev. Lett.}\ }\textbf {\bibinfo {volume} {95}},\
  \bibinfo {pages} {19} (\bibinfo {year} {2005})}\BibitemShut {NoStop}%
\bibitem [{Note4()}]{Note4}%
  \BibitemOpen
  \bibinfo {note} {Note that our notion of prestress is not to be confused with
  the constant prestress externally applied in bulk rheology experiments, which
  is a \protect \emph {shear stress} in general.}\BibitemShut {Stop}%
\bibitem [{Note5()}]{Note5}%
  \BibitemOpen
  \bibinfo {note} {Although we could in principle define total prestress as the
  normal component of the stress acting on \protect \emph {any} plane in our
  system we prefer to use shear planes as this simplifies the forthcoming
  analysis.}\BibitemShut {Stop}%
\bibitem [{\citenamefont {Connelly}\ and\ \citenamefont
  {Back}(1998)}]{Connelly1998a}%
  \BibitemOpen
  \bibfield  {author} {\bibinfo {author} {\bibfnamefont {R.}~\bibnamefont
  {Connelly}}\ and\ \bibinfo {author} {\bibfnamefont {A.}~\bibnamefont
  {Back}},\ }\href {\doibase 10.1511/1998.2.142} {\bibfield  {journal}
  {\bibinfo  {journal} {American Scientist}\ }\textbf {\bibinfo {volume}
  {86}},\ \bibinfo {pages} {142} (\bibinfo {year} {1998})}\BibitemShut
  {NoStop}%
\bibitem [{Note6()}]{Note6}%
  \BibitemOpen
  \bibinfo {note} {Note that individual crosslinks are still under tension;
  however, the total normal force acting on the shear plane
  vanishes.}\BibitemShut {Stop}%
\bibitem [{\citenamefont {Pugh}(1976)}]{pugh1976introduction}%
  \BibitemOpen
  \bibfield  {author} {\bibinfo {author} {\bibfnamefont {A.}~\bibnamefont
  {Pugh}},\ }\href@noop {} {\emph {\bibinfo {title} {An introduction to
  tensegrity}}}\ (\bibinfo  {publisher} {University of California Pr},\
  \bibinfo {year} {1976})\BibitemShut {NoStop}%
\bibitem [{\citenamefont {Alexander}(1998)}]{Alexander1998}%
  \BibitemOpen
  \bibfield  {author} {\bibinfo {author} {\bibfnamefont {S.}~\bibnamefont
  {Alexander}},\ }\href {\doibase 10.1016/S0370-1573(97)00069-0} {\bibfield
  {journal} {\bibinfo  {journal} {Physics Reports}\ }\textbf {\bibinfo {volume}
  {296}},\ \bibinfo {pages} {65} (\bibinfo {year} {1998})}\BibitemShut
  {NoStop}%
\bibitem [{\citenamefont {Heussinger}\ and\ \citenamefont
  {Frey}(2007)}]{Heussinger2007e}%
  \BibitemOpen
  \bibfield  {author} {\bibinfo {author} {\bibfnamefont {C.}~\bibnamefont
  {Heussinger}}\ and\ \bibinfo {author} {\bibfnamefont {E.}~\bibnamefont
  {Frey}},\ }\href
  {http://link.springer.com/article/10.1140/epje/i2007-10209-1} {\bibfield
  {journal} {\bibinfo  {journal} {The European Physical Journal E}\ ,\ \bibinfo
  {pages} {1}} (\bibinfo {year} {2007})},\ \Eprint
  {http://arxiv.org/abs/arXiv:0705.1425v2} {arXiv:arXiv:0705.1425v2}
  \BibitemShut {NoStop}%
\bibitem [{\citenamefont {Huisman}\ \emph {et~al.}(2007)\citenamefont
  {Huisman}, \citenamefont {van Dillen}, \citenamefont {Onck},\ and\
  \citenamefont {{Van der Giessen}}}]{Huisman2007}%
  \BibitemOpen
  \bibfield  {author} {\bibinfo {author} {\bibfnamefont {E.~M.}\ \bibnamefont
  {Huisman}}, \bibinfo {author} {\bibfnamefont {T.}~\bibnamefont {van Dillen}},
  \bibinfo {author} {\bibfnamefont {P.~R.}\ \bibnamefont {Onck}}, \ and\
  \bibinfo {author} {\bibfnamefont {E.}~\bibnamefont {{Van der Giessen}}},\
  }\href {\doibase 10.1103/PhysRevLett.99.208103} {\bibfield  {journal}
  {\bibinfo  {journal} {Phys. Rev. Lett.}\ }\textbf {\bibinfo {volume} {99}},\
  \bibinfo {pages} {2} (\bibinfo {year} {2007})}\BibitemShut {NoStop}%
\bibitem [{\citenamefont {Žagar}\ \emph {et~al.}(2011)\citenamefont
  {Žagar}, \citenamefont {Onck},\ and\ \citenamefont {{Van der
  Giessen}}}]{Zagar2011}%
  \BibitemOpen
  \bibfield  {author} {\bibinfo {author} {\bibfnamefont {G.}~\bibnamefont
  {Žagar}}, \bibinfo {author} {\bibfnamefont {P.~R.}\ \bibnamefont {Onck}}, \
  and\ \bibinfo {author} {\bibfnamefont {E.}~\bibnamefont {{Van der
  Giessen}}},\ }\href {\doibase 10.1021/ma201257v} {\bibfield  {journal}
  {\bibinfo  {journal} {Macromolecules}\ }\textbf {\bibinfo {volume} {44}},\
  \bibinfo {pages} {7026} (\bibinfo {year} {2011})}\BibitemShut {NoStop}%
\bibitem [{\citenamefont {Koenderink}\ \emph {et~al.}(2009)\citenamefont
  {Koenderink}, \citenamefont {Dogic}, \citenamefont {Nakamura}, \citenamefont
  {Bendix}, \citenamefont {MacKintosh}, \citenamefont {Hartwig}, \citenamefont
  {Stossel},\ and\ \citenamefont {Weitz}}]{Koenderink2009}%
  \BibitemOpen
  \bibfield  {author} {\bibinfo {author} {\bibfnamefont {G.~H.}\ \bibnamefont
  {Koenderink}}, \bibinfo {author} {\bibfnamefont {Z.}~\bibnamefont {Dogic}},
  \bibinfo {author} {\bibfnamefont {F.}~\bibnamefont {Nakamura}}, \bibinfo
  {author} {\bibfnamefont {P.~M.}\ \bibnamefont {Bendix}}, \bibinfo {author}
  {\bibfnamefont {F.~C.}\ \bibnamefont {MacKintosh}}, \bibinfo {author}
  {\bibfnamefont {J.~H.}\ \bibnamefont {Hartwig}}, \bibinfo {author}
  {\bibfnamefont {T.~P.}\ \bibnamefont {Stossel}}, \ and\ \bibinfo {author}
  {\bibfnamefont {D.~A.}\ \bibnamefont {Weitz}},\ }\href {\doibase
  10.1073/pnas.0903974106} {\bibfield  {journal} {\bibinfo  {journal} {Proc.
  Natl. Acad. Sci. U.S.A.}\ }\textbf {\bibinfo {volume} {106}},\ \bibinfo
  {pages} {15192} (\bibinfo {year} {2009})}\BibitemShut {NoStop}%
\end{thebibliography}%

\end{document}